\RequirePackage[colorlinks,allcolors=blue]{hyperref}
\documentclass[draft]{agujournal2019}
\usepackage{url} 
\usepackage{lineno}
\usepackage{soul}
\usepackage{aas_macros}
\usepackage{doi}
\usepackage{gensymb}
\def \be {\begin{equation}}
\def \ee {\end{equation}}

\let\olddegree\degree  
\let\degree\relax      
\usepackage{amsmath}
\usepackage{comment}
\usepackage{mathabx}
\let\degree\olddegree  

\draftfalse

\journalname{JGR: Planets}

\begin{document}

%


\title{Geodynamics of super-Earth GJ 486b}

%
%

\authors{Tobias G. Meier\affil{1,2}, Dan J. Bower\affil{2,3}, Tim Lichtenberg\affil{4}, Mark Hammond\affil{1}, Paul J. Tackley\affil{5}, Raymond T. Pierrehumbert\affil{1}, Jos\'e~A. Caballero\affil{6}, Shang-Min Tsai\affil{7}, Megan Weiner Mansfield\affil{8,9}, Nicola Tosi\affil{10}, Philipp Baumeister\affil{10,11}}

\affiliation{1}{Atmospheric, Oceanic and Planetary Physics, Department of Physics, University of Oxford, Parks Road, Oxford OX1 3PU, UK}
\affiliation{2}{Center for Space and Habitability,  University of Bern, Gesellschaftsstrasse 6, 3012 Bern, Switzerland}
\affiliation{3}{Institute of Geochemistry and Petrology, Department of Earth and Planetary Sciences, ETH Zurich, Clausiusstrasse 25, Zurich 8092, Switzerland}
\affiliation{4}{Kapteyn Astronomical Institute, University of Groningen, PO Box 800, 9700 AV Groningen, The Netherlands}
\affiliation{5}{Institute of Geophysics, Department of Earth and Planetary Sciences, ETH Zurich, Sonneggstrasse 5, Zurich 8092, Switzerland}
\affiliation{6}{Centro de Astrobiolog\'ia, CSIC-INTA, ESAC Campus, Camino bajo del castillo s/n, 28049 Villanueva de la Ca\~nada, Madrid, Spain}
\affiliation{7}{Department of Earth and Planetary Sciences, University of California, Riverside, USA}
\affiliation{8}{Steward Observatory, University of Arizona, Tucson, AZ 85719, USA}
\affiliation{9}{NHFP Sagan Fellow}
\affiliation{10}{Institute of Planetary Research, German Aerospace Center (DLR), Germany}
\affiliation{11}{Institut für Geologische Wissenschaften, Freie Universität Berlin, Berlin, Germany}

\correspondingauthor{Tobias G. Meier}{tobias.meier@physics.ox.ac.uk}


\begin{keypoints}
\item The mantle dynamics of super-Earth GJ~486b are governed by the strength of the lithosphere and the day-night surface temperature contrast
\item Degree-1 convection is a consequence of the strong lithosphere, rather than the temperature contrast between the dayside and nightside 
\item A strong surface temperature contrast between the dayside and nightside can anchor downwellings to one hemisphere
\end{keypoints}

%



\begin{abstract}
Many super-Earths are on very short orbits around their host star and, therefore, more likely to be tidally locked. Because this locking can lead to a strong contrast between the dayside and nightside surface temperatures, these super-Earths could exhibit mantle convection patterns and tectonics that could differ significantly from those observed in the present-day solar system. The presence of an atmosphere, however, would allow transport of heat from the dayside towards the nightside and thereby reduce the surface temperature contrast between the two hemispheres. On rocky planets, atmospheric and geodynamic regimes are closely linked, which directly connects the question of atmospheric thickness to the potential interior dynamics of the planet. 
Here, we study the interior dynamics of super-Earth GJ~486b ($R=1.34$\,$R_{\Earth}$, $M=3.0$\,$M_{\Earth}$, T$_\mathrm{eq}\approx700$\,K), which is one of the most suitable M-dwarf super-Earth candidates for retaining an atmosphere produced by degassing from the mantle and magma ocean.  We investigate how the geodynamic regime of GJ~486b is influenced by different surface temperature contrasts by varying possible atmospheric circulation regimes. We also investigate how the strength of the lithosphere affects the convection pattern.
We find that hemispheric tectonics, the surface expression of degree-1 convection with downwellings forming on one hemisphere and upwelling material rising on the opposite hemisphere,  is a consequence of the strong lithosphere rather than surface temperature contrast. Anchored hemispheric tectonics, where downwellings und upwellings have a preferred (day/night) hemisphere, is favoured for strong temperature contrasts between the dayside and nightside and higher surface temperatures.

\end{abstract}

\section*{Plain Language Summary}
The tectonic processes occurring on super-Earths, which are rocky exoplanets with masses exceeding that of Earth, may differ significantly from those observed on the terrestrial planets in our solar system. Many super-Earths are also expected to be tidally locked to their host star, so that always the same hemisphere faces the star. Tidal locking can lead to a strong temperature contrast between the dayside and nightside surface. Here, we investigate the influence of such strong surface temperature contrasts on the interior dynamics and tectonic behaviour of super-Earth GJ~486b, which is one of the best characterised Earth-mass planets to date. We determined the surface temperature contrast between the dayside and nightside by assuming different efficiencies of atmospheric heat transport and ran global climate models for a set of different atmospheric compositions.  
Our 2D mantle convection simulations reveal that hemispheric tectonics, where cold material sinks on one hemisphere and hot material rises on the other side, emerges regardless of the surface temperature contrast if the planet has a strong lithosphere. However, a significant surface temperature contrast and an overall higher surface temperature tend to anchor the sinking cold material and rising hot material to the dayside and nightside respectively.

%
%

\section{Introduction} \label{section:intro}
The tectonic behaviours of super-Earths, characterised as rocky exoplanets with masses exceeding that of Earth, may differ significantly from those of the terrestrial planets in our solar system \cite<e.g.>[]{Lichtenberg2024}. Within our solar system, Earth is the only planet with active plate tectonics. In contrast, Mercury Mars, and the Moon are expected to be in a stagnant lid regime, where the lithosphere is predominantly immobile \cite{Solomatov1997}. While Venus currently lacks plate tectonics, its young surface suggests that it could be in a so-called episodic lid regime \cite<intermittent episodes of mobility -- e.g.>[]{Moresi1998a, Schubert2001, Armann2012}, a plutonic-squishy lid \cite<hot intrusions lead to squishy lithosphere and delamination of crust -- e.g.>[]{Lourenco2020, Smrekar2023}, or a deformable episodic lid \cite<high surface mobility with episodic overturns -->[]{Tian2023a}. 
The terrestrial bodies in our solar system all have sizes that are comparable to or smaller than Earth's, and as such, there is no super-Earth analogue among them. For such massive exoplanets with an irradiation comparable to Earth and with short rotation period, it is still an open question whether plate tectonics is more-likely \cite<e.g.>[]{Valencia2007a, Heck2011} or less likely \cite<e.g.>[]{ONeill2007a, Miyagoshi2015}. However, the majority of known super-Earths is on very short orbits and therefore likely to be tidally locked such that always the same hemisphere faces the host star \cite<e.g.>[]{Peale1977, Barnes2017, Lyu2024}. This locking can lead to strong surface temperature contrasts between the dayside and nightside that have been shown to influence the global convection regime \cite{Gelman2011, VanSummeren2011, Meier2021, Meier2023}. 
          
Here, we investigate the possible tectonic regimes of the (most likely) tidally-locked super-Earth GJ~486b, which has a radius of $1.31$\,$R_{\oplus}$ and a mass of $2.81$\,$M_{\oplus}$ \cite{Trifonov2021a} (see \citeA{Caballero2022} for an alternative determnation of $R$ and $M$). \citeA{Trifonov2021a} placed GJ~486b on the `Earth-like' (50\% enstatite, 50\% iron) composition line in the mass-radius diagram \cite{Trifonov2021a}. Because the planet is on a very short orbit of $1.467$\,days around its red dwarf host star and has a high equilibrium temperature of about $700$\,K, it is also a suitable target for observations that aim to characterise its potential atmosphere \cite{Trifonov2021a}. Observations by the James Webb Space Telescope (JWST) have revealed that GJ~486b could have a water-rich atmosphere, although this feature could also have been caused by contaminating water-rich starspots \cite{Moran2023}. High-resolution spectroscopy observations did not detect a cloud-free atmosphere, although it is still possible that GJ~486b has a high mean molecular weight or H$_2$/He-dominated atmosphere with clouds \cite{RiddenHarper2023}.

It is still debated whether planets on such close orbits around their host-star are able to retain an atmosphere \cite<e.g.>[]{SanzForcada2011, Zahnle2017, Owen2019}. Most super-Earths are likely to lose their secondary atmosphere that was generated from magma ocean outgassing, but depending on the planet's initial volatile budget this atmosphere could potentially be revived through volcanic outgassing \cite{Kite2020, Baumeister2023}. A secondary, predominantly H$_2$ atmosphere detection on the close-in super-Earth exoplanet GJ 1132\,b was suggested by \citeA{Swain2021}. Such an atmosphere would be expected to have been built up through degassing a reduced mantle \cite{Ortenzi2020,Guimond2023a} or magma ocean \cite{Lichtenberg2021b}, but would be short-lived due to intense atmospheric escape. \citeA{Mugnai2021} analysed the same dataset, and their findings suggest, however, that GJ 1132\,b does not have a clear, hydrogen-dominated atmosphere. Follow-up observations favoured the case of no cloud-free atmosphere \cite{LibbyRoberts2022}. To add to the conundrum, \citeA{May2023} next observed two transits of GJ~1132 b with JWST and found that the two transits lead to inconclusive results: one transit is consistent with either an H$_2$O dominated atmosphere or stellar contamination from water-dominated star spots, whereas they found a featureless spectrum for the second transit that ruled out a clear atmosphere. 

The presence of an atmosphere makes heat circulation between the dayside and nightside more efficient, depending on its physical and chemical properties \cite{Pierrehumbert2019,Koll2022}. The corresponding temperature contrast could therefore be less pronounced, which could lead to different tectonic regimes compared to a planet that is devoid of an atmosphere, where the temperature contrast would be the largest. Here, we therefore assume different types of atmospheres for the super-Earth GJ~486b in order to study the possible tectonic regimes that this planet might exhibit. In particular, we investigate whether a strong temperature contrast between the dayside and nightside favours an anchored hemispheric tectonic regime where the downwellings and upwellings have a preferred hemisphere.   

Hemispheric tectonics is the surface expression of a spherical harmonics degree-1 mantle convection pattern (downwellings one one hemisphere and upwellings on the other side). Here, we define anchored hemispheric tectonics as a special case of hemispheric tectonics, where the downwellings get anchored to either the dayside or nightside and the upwellings are preferentially on the other side. Mobile hemispheric tectonics, on the other hand, is associated with a degree-1 convection pattern where the downwellings and upwellings have no preferred hemisphere.   
A degree-1 convection pattern could have existed in the mantles of the Moon \cite<e.g.>[]{Zhong2000}, Mars \cite<e.g.>[]{Zhong2001, Keller2009}, or Earth \cite<e.g.>[]{Zhong2007, McNamara2005}. Degree-1 convection could be induced by depth- and temperature-dependence of viscosity \cite{Tackley1996a, McNamara2005, Yoshida2006}, small core size \cite{Zhong2000}, or if the mantle viscosity is stratified with a weak asthenosphere \cite{Zhong2001, Roberts2006}. An endothermic phase change, such as spinel to bridgmanite and magnesiow\"ustite, is also expected to promote long-wavelength patterns \cite{Tackley1993, Tackley1996a}. On Mars, a transient degree-1 convection pattern could have formed after a giant impact leading to massive volcanism and the formation of a large volcanic province (Tharsis) \cite{Golabek2011, Citron2018}.
Earth's current convection pattern has been predominantly degree-2 for at least the past 120\,Myr as indicated by Earth's subduction history \cite{McNamara2005a, Zhong2007}. 
 
With the discovery of more and more rocky exoplanets, attention has been drawn towards understanding the convection patterns of these distant worlds. \citeA{VanSummeren2011} have shown that for Earth-sized planets, a surface temperature contrast of more than $400$\,K is able to maintain an asymmetric degree-1 mantle convection pattern. Similarly, \citeA{Gelman2011} found that tidally locked planets that receive intense insolation from their host star are more likely to develop a degree-1 convection pattern with mantle upwellings preferentially located at the substellar point. Such a convection pattern could lead to melt production and volcanic outgassing that occurs preferentially on one hemisphere, potentially creating a crustal dichotomy similar to that of Mars \cite<e.g.>[]{Cheng2024}. Such a dichotomy could lead to different reflectance and emission properties between the dayside and nightside surface. If volcanic outgassing occurs predominantly on one side, it would likely influence the planet's atmospheric composition and evolution. Understanding the diversity of tectonics that rocky super-Earths exhibit is therefore essential for understanding current and future observations of rocky super-Earths and assess their potential to host atmospheres. 

\citeA{Meier2021} analysed the possible tectonic regimes of super-Earth LHS\,3844b. This planet has a radius of 1.3\,$R_{\oplus}$ and is most likely devoid of an atmosphere, and its thermal phase curve observation suggest a strong surface temperature contrast with a dayside temperature slightly above $1000$\,K at the substellar point and $0$--$700$\,K on the nightside \cite{Kreidberg2019}. \citeA{Meier2021} suggested that this planet might exhibit hemispheric tectonics with a strong downwelling on the dayside and upwellings that are getting pushed towards the nightside in the case of a strong lithosphere. If the planet has a weak lithosphere, a uniform tectonic regime is established with downwellings and upwellings uniformly distributed around the mantle. 
  
In this study, we therefore examine the effect of lithospheric strength on the degree of convection of super-Earth GJ~486b. Compared to the study of LHS\,3844b by \citeA{Meier2021}, who modelled the interior dynamics assuming no atmospheric heat redistribution, the focus of this new study lies on investigating the impact of atmospheric heat circulation. Specifically, we explore how the convection patterns are affected by varying surface temperature contrasts between the dayside and nightside. 
In particular, we show that degree-1 convection results from a strong lithosphere while anchored hemispheric tectonics is a consequence of both a strong lithosphere and a strong temperature contrast and higher surface temperatures. 

\section{Methods} \label{sec:model}
Since it is uncertain as of yet whether super-Earth GJ~486b has retained an atmosphere, we employed general circulation models (GCMs) to estimate the surface temperatures for reasonable end-member atmospheric regimes (see Section~\ref{sect:surface_temperature}). An upper limit for the temperature at the core-mantle-boundary (CMB) can be estimated from the melting temperature of MgSiO$_3$ assuming that the lower mantle is not molten \cite{Fei2021, Stixrude2014, Nomura2014}. As we are interested in statistically steady flows, we neglect the effects of core cooling and therefore set the CMB to be isothermal at a lower estimate of $T_{\text{CMB}}=4500$\,K. We do not include any interior heat sources, so the models are therefore purely basally heated. The effect of adding internal heat sources has been discussed by \citeA{Meier2021}. 
Both mass and radius have been determined for this super-Earth \cite{Trifonov2021a}, and we can therefore calculate the thickness of the mantle and the core size assuming an Earth-like composition. To do this, we solved the internal structure equations
  \begin{subequations} \label{eq:internal_structure}
    \begin{align}
      \frac{dP}{dr} = - \rho(r)g(r) \label{eq:is1}, \\ 
      \frac{dm}{dr} = 4\pi r^{2} \rho(r) \label{eq:is2}, \\ 
      g(r) =  \frac{Gm(r)}{r^2}     \label{eq:is3},
    \end{align}
  \end{subequations}
where $r$ is the radius, $P$ the pressure, $m$ the mass, $G=6.67430 \cdot 10^{-11}$\,m$^{3}$kg$^{-1}$s$^{-2}$  the gravitational constant, $g$ the gravitational acceleration, and $\rho$ the density. 
We determine the density profile $\rho(P)$ as a function of $P$ using a 3\textsuperscript{rd}-order Birch-Murnaghan equation of state \cite{Murnaghan1944, Birch1947}. The corresponding parameters are shown in Table~\ref{tab:density_refstate}.

\begin{table}
\centering 
\caption{Birch-Murnaghan parameters for reference density profiles. The values for the mantle are from \cite{Tackley2013} and the values for the outer core from \cite{Zeng2014}. $\rho_s$ corresponds to the (extrapolated) density at the surface ($P=0$).} 
\begin{tabular}{@{}lllll}
\hline
Mineralogy & $K_{a0}$ (GPa) &  $K_{a0}^{\prime}$ & $\rho_{s}$ (kg/m$^3$) \\
\hline
upper mantle (olivine)   &   163 & 4.0 &  3240 \\ 
transition zone &   85  & 4.0 &  3226  \\ 
bridgmanite      &   210 & 3.9 & 3870  \\ 
post-perovskite &   210 & 3.9 & 3906  \\ 
(outer) core            &   201    & 4.0 & 7050 \\
\hline
\end{tabular}

\label{tab:density_refstate}
\end{table}
Equations~(\ref{eq:internal_structure}a--c) were solved iteratively to find a solution that agrees with the total mass of the planet from observations. The corresponding solution for the radius of the core mantle boundary is $r_{\text{cmb}}\approx 4900$\,km and the mantle depth is $d_{\text{mantle}}\approx 3380$\,km. The surface gravitational acceleration is $g=16.4$\,m\,s$^{-2}$, which is assumed constant throughout the mantle as it varies by less than $2$\,m\,s$^{-2}$ from the surface to the CMB.

We modelled the mantle convection of GJ~486b using the code StagYY \cite{Tackley2008} in a two-dimensional (2D) spherical annulus geometry \cite{Hernlund2008} using dimensional units. Viscous forces dominate interior convection of rocky planets, and we can therefore make the infinite Prandtl number approximation (i.e. inertial forces are neglected). Compressibility is modelled by employing the truncated anelastic liquid approximation (TALA) where a reference state density profile is prescribed that varies with depth. The density profile is calculated using a 3\textsuperscript{rd}-order Birch-Murnaghan equation of state with the parameters given in Table~\ref{tab:density_refstate}. Material properties, such as thermal expansivity $\alpha$ and thermal conductivity $k$ are also pressure-dependent using the same parameters as in \citeA{Tackley2013}. 
$\alpha$ decreases from its surface value $3\cdot10^{-5}$\,K$^{-1}$ to around $5\cdot10^{-6}$\,K$^{-1}$ at the CMB, and k increases from $3$\,Wm$^{-1}$K$^{-1}$ at the surface to $10$\,Wm$^{-1}$K$^{-1}$ at the CMB.   
All models have a resolution of $256$ cells in the angular direction and $128$ cells in the radial direction which corresponds to a mean spacing of around $1.4^{\circ}$ in the angular direction and $26$\,km in the radial direction. The initial temperature field has a a potential temperature of $2000$\,K and thermal boundary layers at the surface and CMB of thickness $d_{\mathrm{TBL}}=260$\,km. 
The mantle is divided into $3$ viscosity layers: upper mantle, bridgmanite, and post-perovskite layer. Viscosity depends on pressure and temperature and is determined using an Arrhenius-like law
\be \label{eq:chap3_arrhenius}
\eta(P,T) = \eta_0\exp{\left(\frac{E_a+PV_a(P)}{RT}-\frac{E_a}{RT_0}\right)}, 
\ee 
where $\eta_0$ is a reference viscosity,  $E_a$ is the activation energy, $V_a(P)$ is the activation volume, and $R=8.311$\,JK$^{-1}$mol$^{-1}$ is the universal gas constant. The activation volume depends on pressure and is given by:
\be \label{eq:activation_volume}
V_a(P) = V_0 \exp{\left(-\frac{P}{P_{\mathrm{decay}}}\right)},
\ee
where $P_{\mathrm{decay}}$ is the decay pressure controlling the pressure dependence of the activation volume \cite{Karato1993, Tackley2013}. Table~\ref{tab:eta_pars} shows the parameters used for the viscosity for each layer \cite{Tackley2013}. We use a minimum and maximum cutoff for the viscosity of $\eta_{\mathrm{min}}=10^{18}$\,Pa$\cdot$s and $\eta_{\mathrm{max}}=10^{28}$\,Pa$\cdot$ s.

\begin{table*}\centering 
\small
\centering
\caption{From \cite{Tackley2013}: Parameters used for the Arrhenius-like viscosity law (Eqs.~\ref{eq:chap3_arrhenius} and \ref{eq:activation_volume}).}
\begin{tabular}{@{}lcccccccc}\hline
Mineralogy & $E_a$ (kJ/mol) & $V_0$ (cm$^3$/mol) & $P_{\mathrm{decay}}$ (GPa) & $\eta_0$ (Pa$\cdot$s) \\ 
\hline
Upper mantle & $300$ & $5.0$ & $\infty$ & $10^{21}$ \\ 
Bridgmanite & $370$ & $3.65$ & $200$ & $3.0 \cdot 10^{23}$ \\ 
Post-perovskite lower bound & $162$ & $1.4$ & $1610$ & $1.9 \cdot 10^{21}$ \\ 
\hline
\end{tabular}
\label{tab:eta_pars}
\end{table*}
We also model the strength of the lithosphere (outermost rigid layer of the planet) through a plastic yielding criterion. At low pressure, this strength is related to its fracture strength of frictional sliding of faults (Byerlee's law \cite{Byerlee1978}). At high pressure, the strength is related to ductile failure caused by dislocation motion of the lattice \cite{Kohlstedt1995}. Both the brittle and ductile components are encapsulated within a pressure-dependent yield stress $\sigma_{y}$
\be 
\sigma_{\text{y}} = \text{min}(c+c_fP,\sigma_{\text{duct}}+\sigma^{\prime}_{\text{duct}}P),
\ee  
where $c$ is the cohesive strength, $c_f$ is the friction coefficient, $\sigma_{\text{duct}}$ is the ductile yield strength, and $\sigma^{\prime}_{\text{duct}}$ is the ductile yield stress gradient. 
If the stress exceeds the yield stress $\sigma_{y}$, the viscosity gets reduced to an effective viscosity given by  
\be \label{chap3:eta_eff}
\eta_{\text{eff}} = \frac{\sigma_{\text{y}}}{2 \dot{\epsilon_{\text{II}}}} \quad \mathrm{if} \hspace{0.5em} 2 \eta \epsilon_{\text{II}} > \sigma_y,
\ee 
where $\dot{\epsilon_{\text{II}}}$ is the second invariant of the strain rate tensor and $\eta$ is calculated using Equation~\ref{eq:chap3_arrhenius}.  
From laboratory experiments, the strength of the lithosphere is on the order of several hundred MPa \cite{Kohlstedt1995}, while numerical simulations often use a ductile yield strength smaller than $200$\,MPa in order to obtain a plate-like style of convection (mobile lid) \cite<e.g.>[]{Moresi1998a, Richards2001, Lourenco2020}.   
In this study,  we use $c=1$\,MPa, $c_f = 1$, and $\sigma_{\text{duct}}$ is varied from $10$ (weak lithosphere) to $300$\,MPa (strong lithosphere). For the ductile yield stress gradient, we use $\sigma^{\prime}_{\text{duct}}=0.01$ in order to avoid yielding of the deep mantle \cite{Tackley2013}. The brittle components are chosen so that the uppermost cells encompass the brittle behaviour with a yield stress that is $1$ to $2$ orders of magnitude lower than the ductile component at higher pressures. These values are calibrated to give plate tectonic-like behaviour on an Earth-size planet \cite<e.g.>[]{Moresi1998a, Tackley2000, Tackley2013} and represent the large-scale-long-term lithospheric strength, including various weakening mechanisms as discussed in \citeA{Gerya2024}. 

\subsection{Tracking of Plumes and Downwellings} \label{sec:tracking}
We ran the models for several overturn timescales (time scale for material to travel through the mantle), so that they reach steady-state and that we can draw statistically meaningful conclusions about the location of upwellings and downwellings in the mantle. This also ensures that any influence of the initial condition is eliminated. We ran most models up to $50$\,Gyr because the higher viscosity leads to more sluggish convection with an overturn timescales of several Gyrs. 
To quantify the mobility and position of plumes and downwellings, we constructed a plume detection and tracking algorithm. We adopt a similar approach as in \citeA{Labrosse2002} where plumes and downwellings are identified using a thermal threshold that depends on depth: For each point of the temperature field, the threshold determines if it is part of a plume, a downwelling or neither. The plumes and downwellings can then be connected using a connected component analysis or using a k-means clustering analysis. Here, we identify plumes and downwellings only as a function of longitude: For each longitude bin, we determine the number of temperature points that are below (downwelling) or above (plume) the temperature threshold. The plumes and downwellings are then identified from the resulting histogram. The width of the bins is chosen depending on the number of longitudinal grid points. A width that is too small may lead to splitting of plumes/downwellings whereas a too large width might lead to a non-detection.

\subsection{Surface Temperature} \label{sect:surface_temperature}

We ran a suite of models with different surface temperatures that are either calculated using theoretical calculations or motivated by GCMs. 
If the planet is tidally locked with no atmosphere, the temperature at the substellar point (denoted by $T_{\mathrm{day}}$)  is given by \cite{Mendez2017, Caballero2022}
\be \label{eq:heat_redistribution}
T_{\mathrm{day}}^4 = \frac{L_{\star}(1-A)}{4 \pi \sigma d^2},
\ee 
where $L_{\star}$ is the luminosity of the host star GJ~486, $R_{\star}$ its radius, $A$ the planet's Bond albedo, $d$ the separation of the planet from the star, $\sigma$ the Stefan-Boltzmann constant. The host star GJ\,486 has an effective temperature of $3340 \pm 54$\,K \cite{Trifonov2021a}, which leads to a substellar point temperature of $T_{\mathrm{day}} \approx833$\,K.
If the planet has an atmosphere that redistributes heat very efficiently, the uniform surface temperature can be estimated from its equilibrium temperature ($T_{\mathrm{eq}}\approx 700$\,K) assuming that there is no greenhouse warming. 
We also ran an averaged heat redistribution model, for which we averaged the dayside and nightside temperature of the models with no atmosphere and the model with efficient heat redistribution.  Additionally, we ran an Earth-like case where we assumed a uniform surface temperature of $300$\,K.

\begin{figure}
    \centering
    \includegraphics[width=\linewidth]{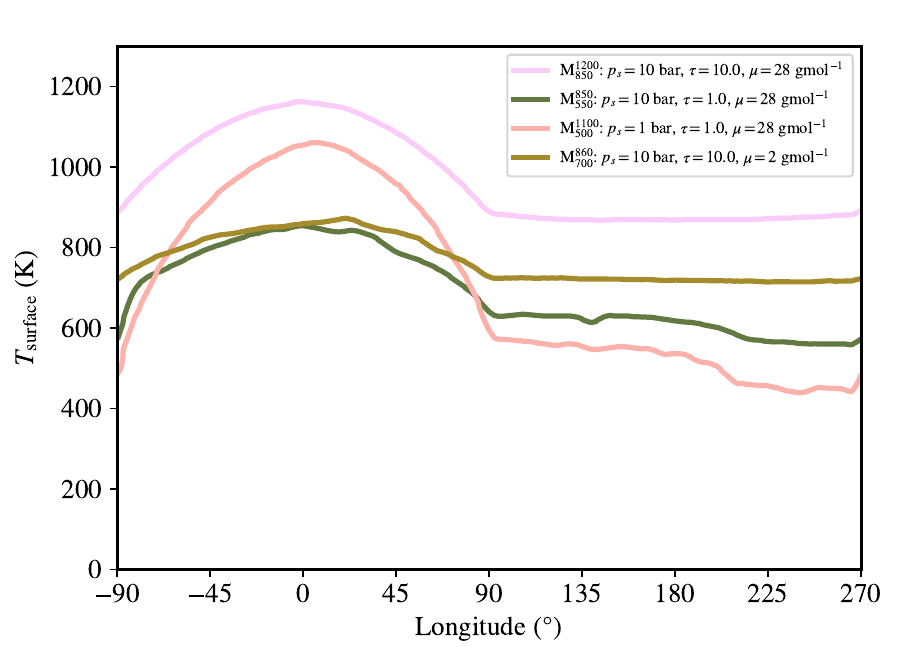}
    \caption{Surface temperatures from GCMs for different surface pressures $p_s$, optical depths $\tau$ and mean molecular weights $\mu$. The corresponding surface temperatures that are used in the models are shown in Figure~\ref{fig:tsurf_gliese_all}.}
    \label{fig:phasecurve_gcm_all}
\end{figure}

Figure \ref{fig:phasecurve_gcm_all} shows the surface temperature from the Exo-FMS GCM simulations \cite{Hammond2017, Hammond2021} for different surface pressures $p_s$, longwave optical depths $\tau$ and mean molecular weights $\mu$. We use the same numerical setup as in \citeA{Hammond2021}, in which Exo-FMS solves the primitive equations on a cubed-sphere grid with a vertical sigma pressure coordinate, using the cubed-sphere dynamical core of the GFDL FMS \cite{Lin2004}. We consider a dry atmosphere with no condensable species. We use an idealised semi-grey radiative transfer scheme to avoid making explicit assumptions about the composition of atmospheric absorbers. This simplified setup lets us explore the effect of a wide range of types of atmosphere on the surface temperature distribution.

We vary the surface pressure $p_s$ between 1 and 10 bar to explore the effect of atmospheric thickness and total heat capacity on the atmospheric heat redistribution. The upper limit of $10$ bar corresponds to an atmosphere that has experienced minimal atmospheric escape since its primordial capture from the solar nebula, while 1 bar is Earth-like, reflecting the assumption that atmospheres with pressures below this threshold would not redistribute heat efficiently. We also vary the longwave optical depth $\tau$ between 1 and 10 to explore the effect of greenhouse warming on the surface temperatures. Finally, we vary the mean molecular weight $\mu$ between $2$  and $28$\,g/mol to explore the effect of atmospheric scale height and heat capacity on the redistribution of heat. A mean molecular weight of $\mu=2$\,g/mol corresponds to a hydrogen-dominated atmosphere, similar to a volatile-rich primordial atmosphere. A mean molecular weight of $\mu=28$\,g/mol corresponds to a nitrogen-dominated atmosphere, akin to a secondary atmosphere that could have formed after the planet has lost its primordial atmosphere. 

Testing each two options for each parameter results in eight simulations in total. From this suite of GCMs, we selected four models which exhibit the most distinct temperature contrasts and determined the corresponding dayside and nightside temperatures for the interior models. The parameters for the GCM models and the corresponding dayside and nightside temperature are shown in Table~\ref{tab:model_pars_gcm}. Atmospheres with a greater optical depth exhibit higher surface temperatures, while a lower mean molecular weight increases heat redistribution, resulting in a less pronounced surface temperature contrast \cite{Hammond2017}. 

\begin{figure}
    \centering
    \includegraphics[width=\linewidth]{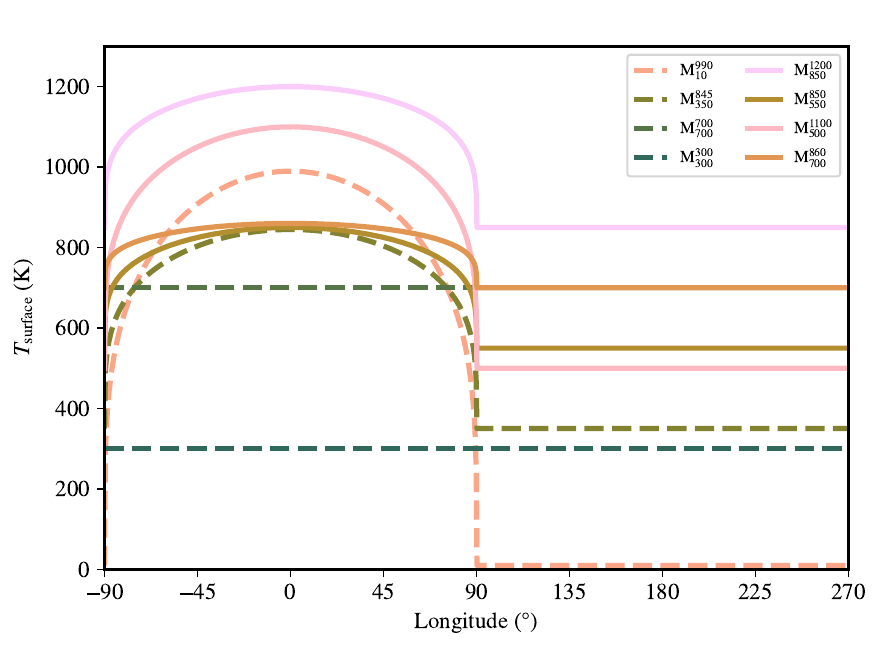}
    \caption{Surface temperatures used for the different model runs. The dashed lines show the models where the temperature contrast has been determined assuming different efficiencies of heat redistribution (Equation~\ref{eq:heat_redistribution}). The solid lines show models where the temperature contrast is inferred from GCM simulations (Figure~\ref{fig:phasecurve_gcm_all}).} 
    \label{fig:tsurf_gliese_all}
\end{figure}

Finally, we use the substellar point and nightside temperatures obtained from both the GCMs output and the theoretical calculations to approximate the longitudinally varying surface temperature boundary condition using
\be \label{eq:tsurf_models}
    T(\theta)= 
\begin{cases}
    T_{\mathrm{night}} + (T_{\mathrm{day}}-T_{\mathrm{night}}) \cos^{1/4}{(\theta)}& \text{if } \theta \in [-90,90) \\
    T_{\mathrm{night}}              & \text{if } \theta \in [90,270).
\end{cases}
\ee 
Table~\ref{tab:model_pars_eff} shows an overview of the different model parameters for the cases with different efficiencies of heat redistribution. We refer to specific model runs using the notation M$^{T_{\text{day}}}_{T_{\text{night}}}-\sigma_{\text{duct}}$. The corresponding surface temperatures used as a thermal surface boundary condition in the interior models are shown in Figure~\ref{fig:tsurf_gliese_all}. Surface temperatures that have been determined assuming different efficiencies of heat redistribution (using Equation~\eqref{eq:heat_redistribution}) are shown with a dashed line. The surface temperatures that have been inferred from GCM simulations are shown with a solid line. 
For the M$^{1200}_{850}$ and M$^{300}_{300}$ model, we also vary the plastic yield strength $\sigma_{\text{duct}}$ from $50$ to $200$\,MPa. The corresponding parameters for the interior models are shown in Tables~\ref{tab:models_yield} and \ref{tab:models_yield_300}. 

\begin{table*}
\caption[Surface temperature contrasts from GCMs]{Dayside and nightside temperatures derived from GCMs with different atmospheric parameters.}
\label{tab:model_pars_gcm}
\centering
\small
\begin{tabular}{ccccccc}
\hline\hline
Model & dayside temp. & nightside temp. & $\sigma_{\text{duct}}$ & \multicolumn{3}{c}{atmospheric parameters} \\ 
&  &   &  & $p_s$ & $\tau$ & $\mu$ \\
& (K) & (K)  & (MPa) & (bar) & & (g/mol) \\
\hline
M$^{1200}_{850}-$100  &   1200  & 850 & 100  & $10$ & $10.0$ & $28$ \\ 
M$^{850}_{550}-$100   &   850   & 550 & 100  & $10$ & $1.0$  & $28$ \\
M$^{1100}_{500}-$100  &   1100  & 500 & 100  & $1$  & $1.0$  & $28$ \\ 
M$^{860}_{700}-$100   &   860   & 700 & 100  & $10$ & $10.0$ & $2$  \\ 
M$^{1200}_{850}-$300  &   1200  & 850 & 300  & $10$  & $10.0$ & $28$ \\ 
M$^{850}_{550}-$300   &   850   & 550 & 300  & $1$  & $1.0$  & $28$ \\
M$^{1100}_{500}-$300  &   1100  & 500 & 300  & $1$  & $10.0$  & $28$ \\ 
M$^{860}_{700}-$300   &   860   & 700 & 300  & $10$ & $10.0$ & $2$  \\ 
\hline
\end{tabular}
\end{table*}

\begin{table*}
\caption{Dayside and nightside temperatures for different efficiencies of heat redistribution. M$^{990}_{10}$ and M$^{700}_{700}$ are deduced from Equation~\ref{eq:heat_redistribution}. M$^{845}_{350}$ is an average between  M$^{990}_{10}$ and M$^{700}_{700}$, representing a case with average heat redistribution. M$^{300}_{300}$ corresponds to a model with Earth's surface temperature.}
\label{tab:model_pars_eff}
\centering
\begin{tabular}{ccccc}
\hline\hline
Model & dayside temp. & nightside temp. & $\sigma_{\text{duct}}$ &heat redistribution \\ 
& (K) & (K) & (MPa) & \\
\hline
M$^{990}_{10}$-300 &   990  & 10  & 300 & tidally locked  \\ 
M$^{845}_{350}$-300 &   845 & 350 & 300 & average heat redistribution \\ 
M$^{700}_{700}$-300 &   700 & 700 & 300 & uniform  \\ 
M$^{300}_{300}$-300 &   300 & 300 & 300 & uniform Earth temp. \\ 
M$^{990}_{10}$-10 &   990  & 10  & 10  & tidally locked   \\ 
M$^{845}_{350}$-10 &   845 & 350 & 10  & average heat redistribution \\ 
M$^{700}_{700}$-10 &   700 & 700 & 10  & uniform  \\ 
M$^{300}_{300}$-10 &   300 & 300 & 10  & uniform Earth temp. \\ 
\hline
\end{tabular}
\end{table*}

\begin{table}
\caption[Parameters for models with variable yield stress (M$^{1200}_{850}$)]{Parameters for the models where the plastic yielding strength is varied for the M$^{1200}_{850}$ model.}	
\label{tab:models_yield}\centering
\begin{tabular}{cccc}
\hline\hline
Model & dayside temp. & nightside temp. & $\sigma_{\text{duct}}$  \\ 
& (K) & (K)  & (MPa) \\
\hline 
M$^{1200}_{850}-$25  & 1200  & 850  & 25 \\ 
M$^{1200}_{850}-$50  & 1200  & 850  & 50 \\ 
M$^{1200}_{850}-$75  & 1200  & 850  & 75 \\ 
M$^{1200}_{850}-$125 & 1200  & 850  & 125 \\
M$^{1200}_{850}-$150 & 1200  & 850  & 150 \\ 
M$^{1200}_{850}-$175 & 1200  & 850  & 175 \\ 
M$^{1200}_{850}-$200 & 1200  & 850  & 200 \\
M$^{1200}_{850}-$225 & 1200  & 850  & 225 \\
M$^{1200}_{850}-$250 & 1200  & 850  & 250 \\
\hline
\end{tabular}
\end{table}

\begin{table}
\caption[Parameters for models with variable yield stress (M$^{300}_{300}$)]{Parameters for the models where the plastic yielding strength is varied for the M$^{300}_{300}$ model.}	
\label{tab:models_yield_300}
\centering
\begin{tabular}{cccc}
\hline\hline
Model & dayside temp. & nightside temp. & $\sigma_{\text{duct}}$  \\ 
& (K) & (K)  & (MPa) \\
\hline
M$^{300}_{300}-$25  & 300  & 300  & 25 \\ 
M$^{300}_{300}-$50  & 300  & 300   & 50 \\ 
M$^{300}_{300}-$75  & 300  & 300   & 75 \\ 
M$^{300}_{300}-$125 & 300  & 300   & 125 \\
M$^{300}_{300}-$150 & 300  & 300   & 150 \\ 
M$^{300}_{300}-$175 & 300  & 300   & 175 \\ 
M$^{300}_{300}-$200 & 300  & 300   & 200 \\
M$^{300}_{300}-$225 & 300  & 300   & 225 \\
M$^{300}_{300}-$250 & 300  & 300   & 250 \\
\hline
\end{tabular}
\end{table}

\section{Results} \label{sec:res}
\subsection{Models with Low Yield Stress ($\sigma_{\text{duct}}=10$\,MPa)}
For the models with a low yield stress ($\sigma_{\text{duct}}=10$\,MPa) (see Table~\ref{tab:model_pars_eff}), we determined the surface temperature contrasts assuming different efficiencies of heat redistribution (Fig.~\ref{fig:tsurf_gliese_all}).  
Figure~\ref{fig:gj_tplot_theo_weak} shows snapshots of the mantle temperature at $50$\,Gyr and the corresponding evolutionary tracks of upwellings and downwellings (see also Section~\ref{sec:tracking}). All models have a uniform distribution of hot upwellings and cold downwellings regardless of the temperature contrast. However, the downwellings are stronger (i.e. colder and more viscous) for the models with a stronger temperature contrast between the dayside and the nightside.  	
As it can be seen from the evolutionary tracks, the location of the upwellings and downwellings is very stable. Downwellings for the M$^{990}_{10}$-10  are not showing up on the evolutionary track plot because they are too faint compared to the downwellings on the nightside to be detected by the downwelling detection algorithm. 
Figures~\ref{fig:gj_tplot_theo_weak}(C, D) show the results for a uniform surface temperature of $700$ and $300$\,K respectively. Downwellings and upwellings are both uniformly distributed and there is (as expected for uniform surface temperatures) no dichotomy between the dayside and nightside. Both downwellings and upwellings are more stable for the $300$\,K case because the lower temperature makes the downwellings more viscous and therefore also more sluggish.   

\subsection{Models with Medium Yield Stress ($\sigma_{\text{duct}}=100$\,MPa)}
\begin{figure*}
    \centering
    \includegraphics[width=\linewidth]{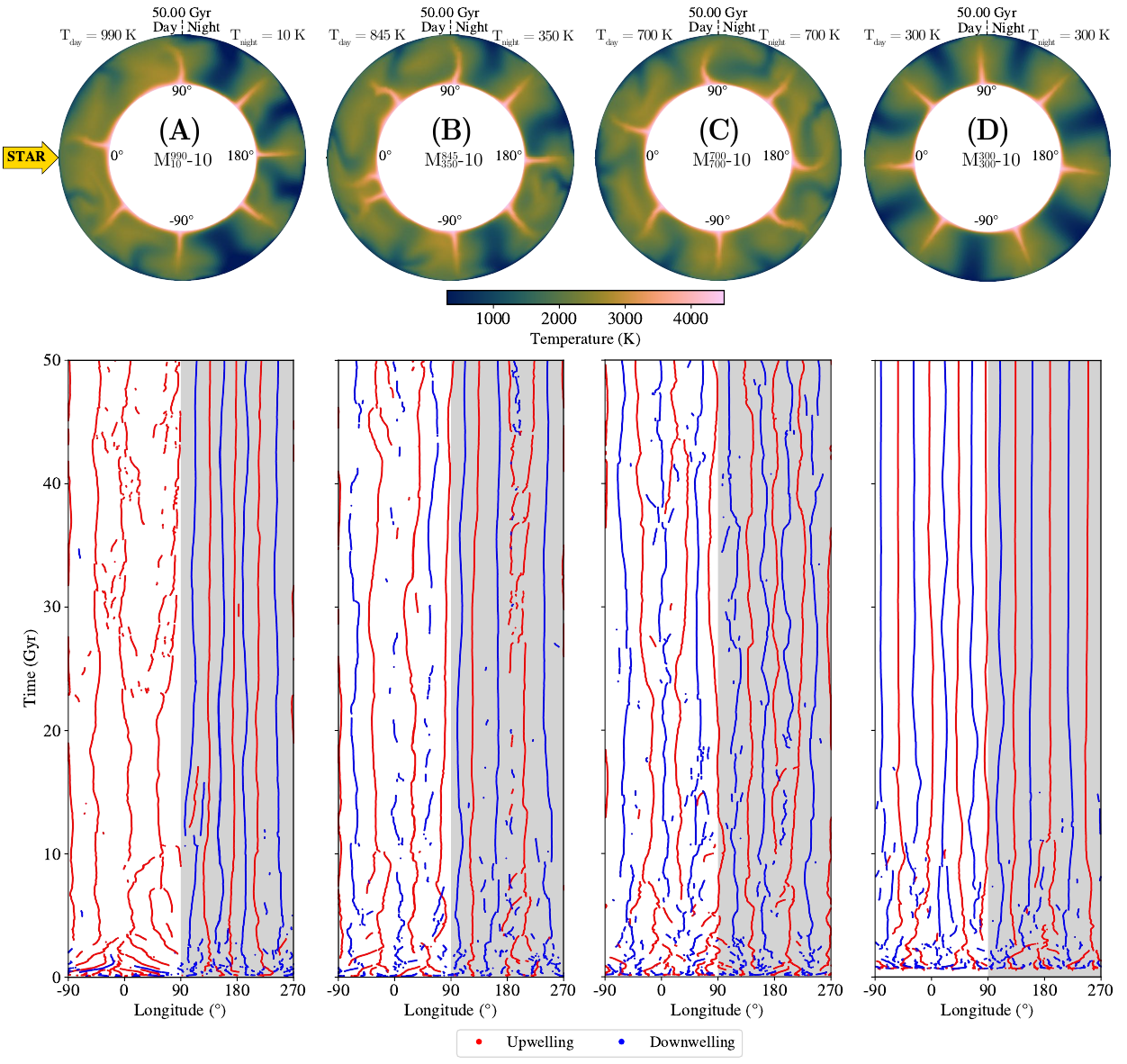}
    \caption{Snapshots of mantle temperature for super-Earth GJ~486b at $50$\,Gyr for the models with a weak lithosphere (\mbox{$\sigma_{\text{duct}}=10$\,MPa}) (top row) and the corresponding evolutionary tracks of upwellings (red) and downwellings (blue) (bottom row). The substellar point is at $0\degree$ longitude (indicated by the yellow arrow) and the nightside ($90\degree$--$270\degree$) is denoted by a grey background. Each column (A-D) corresponds to a different surface temperature contrast.} 
    \label{fig:gj_tplot_theo_weak}
\end{figure*}

\begin{figure*}
    \centering
    \includegraphics[width=\linewidth]{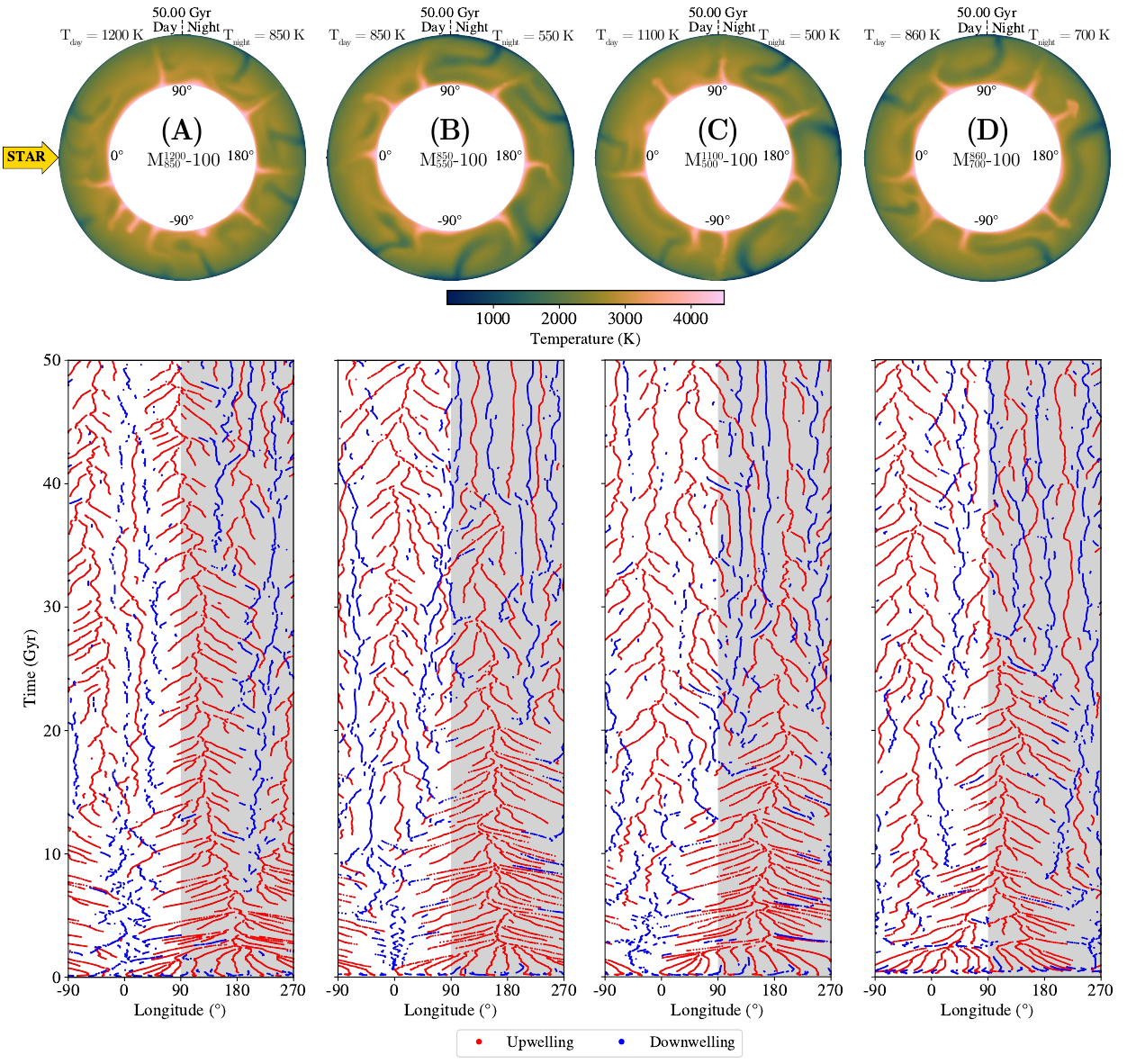}
    \caption{Same as Figure~\ref{fig:gj_tplot_theo_weak} but for moderately strong lithosphere (\mbox{$\sigma_{\text{duct}}=100$\,MPa}) and with surface temperatures that are derived from general circulation models. The substellar point is at $0\degree$ longitude (indicated by the yellow arrow) and the nightside ($90\degree$--$270\degree$) is denoted by a grey background. Each column (A-D) corresponds to a different surface temperature contrast. Compared to the models with a low yield stress (Fig.~\ref{fig:gj_tplot_theo_weak}), downwellings are now thinner, especially for high dayside temperatures.}
    \label{fig:gj_gcm_tzebra_medium}
\end{figure*}

Figure~\ref{fig:gj_gcm_tzebra_medium} shows snapshots of the mantle temperature and evolutionary tracks of upwellings and downwellings for the models with a medium yield stress ($\sigma_{\text{duct}}=100$\,MPa). The surface temperature contrasts for these models have been derived using GCMs. 
Upwellings are uniformly distributed for all cases. For high dayside surface temperatures (models M$^{1200}_{850}$-100, M$^{1100}_{500}$-100), downwellings are thin on the dayside and diffuse away quickly (Fig.~\ref{fig:gj_gcm_tzebra_medium}: A, C). For lower dayside surface temperatures (models M$^{850}_{550}$-100, M$^{860}_{700}$-100), downwellings also develop on the dayside and are able to subduct into the deep mantle (Fig.~\ref{fig:gj_gcm_tzebra_medium}: B,D). Several cold and viscous downwellings develop on the nightside for all models, but they are less strong as for the models with a low yield stress. 
By comparing the evolutionary tracks with those of the models featuring a weak lithosphere (Fig.~\ref{fig:gj_tplot_theo_weak}), it becomes apparent that the downwellings and upwellings exhibit greater mobility in the case with $\sigma_{\text{duct}}=100$\,MPa. Due to the reduced viscosity of the downwellings, they cannot effectively anchor the plumes in place, resulting in increased time dependence. 

\subsection{Models with High Yield Stress ($\sigma_{\text{duct}}=300$\,MPa)}
\begin{figure*}
    \centering
    \includegraphics[width=\linewidth]{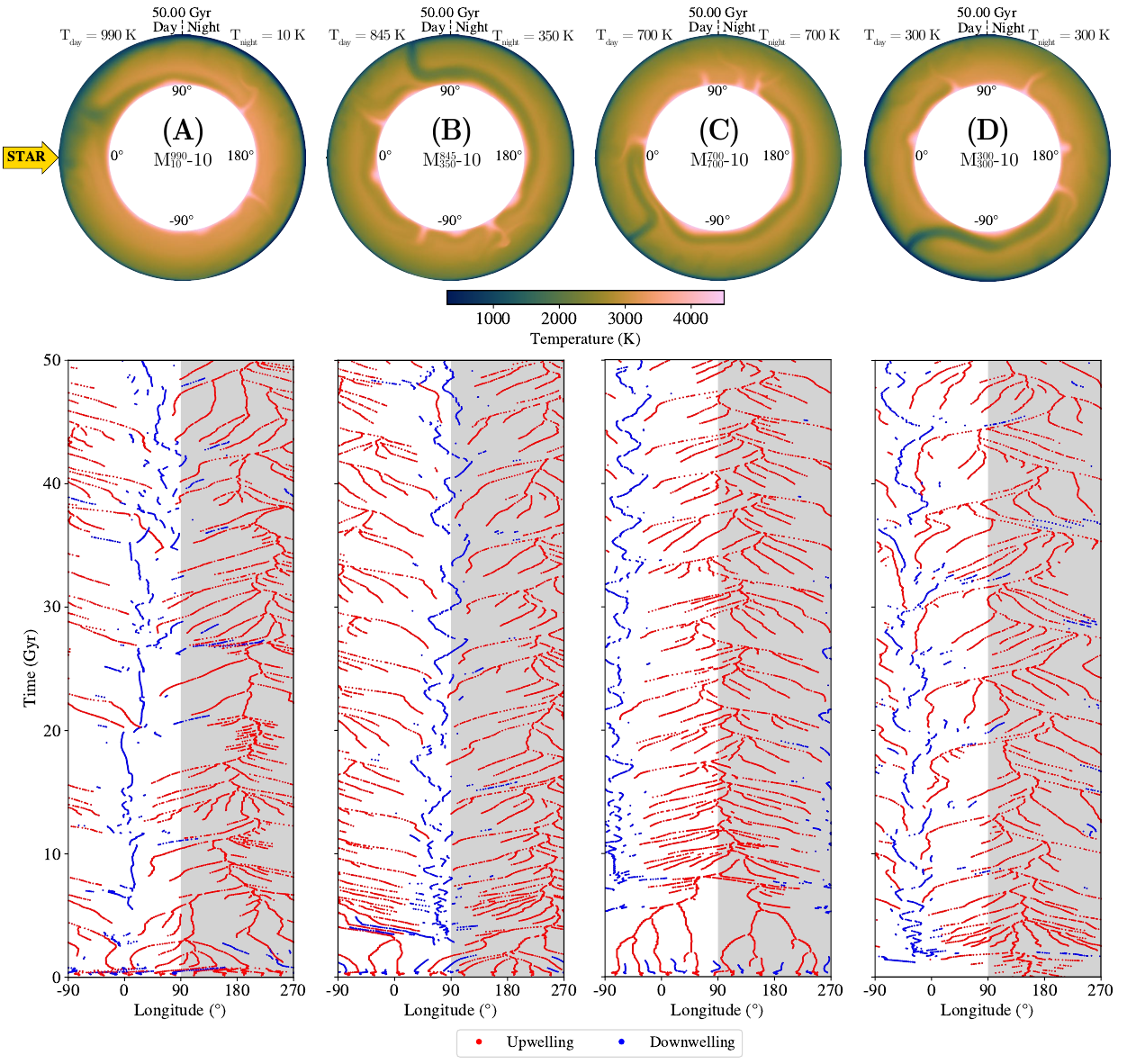}
    \caption{Same as Figure~\ref{fig:gj_tplot_theo_weak} but for the models with a strong lithosphere (\mbox{$\sigma_{\text{duct}}=300$\,MPa}). Each column (A-D) corresponds to a different surface temperature contrast. Surface temperature contrasts are derived assuming different efficiencies of heat redistribution. The substellar point is at $0\degree$ longitude (indicated by the yellow arrow) and the nightside ($90\degree$--$270\degree$) is denoted by a grey background. A degree-1 convection pattern is established for all models with a strong downwelling forming on one hemisphere and upwellings that are being pushed to the other side.}
    \label{fig:gj_theo_tzebra}
\end{figure*}

\begin{figure*}
    \centering
    \includegraphics[width=\linewidth]{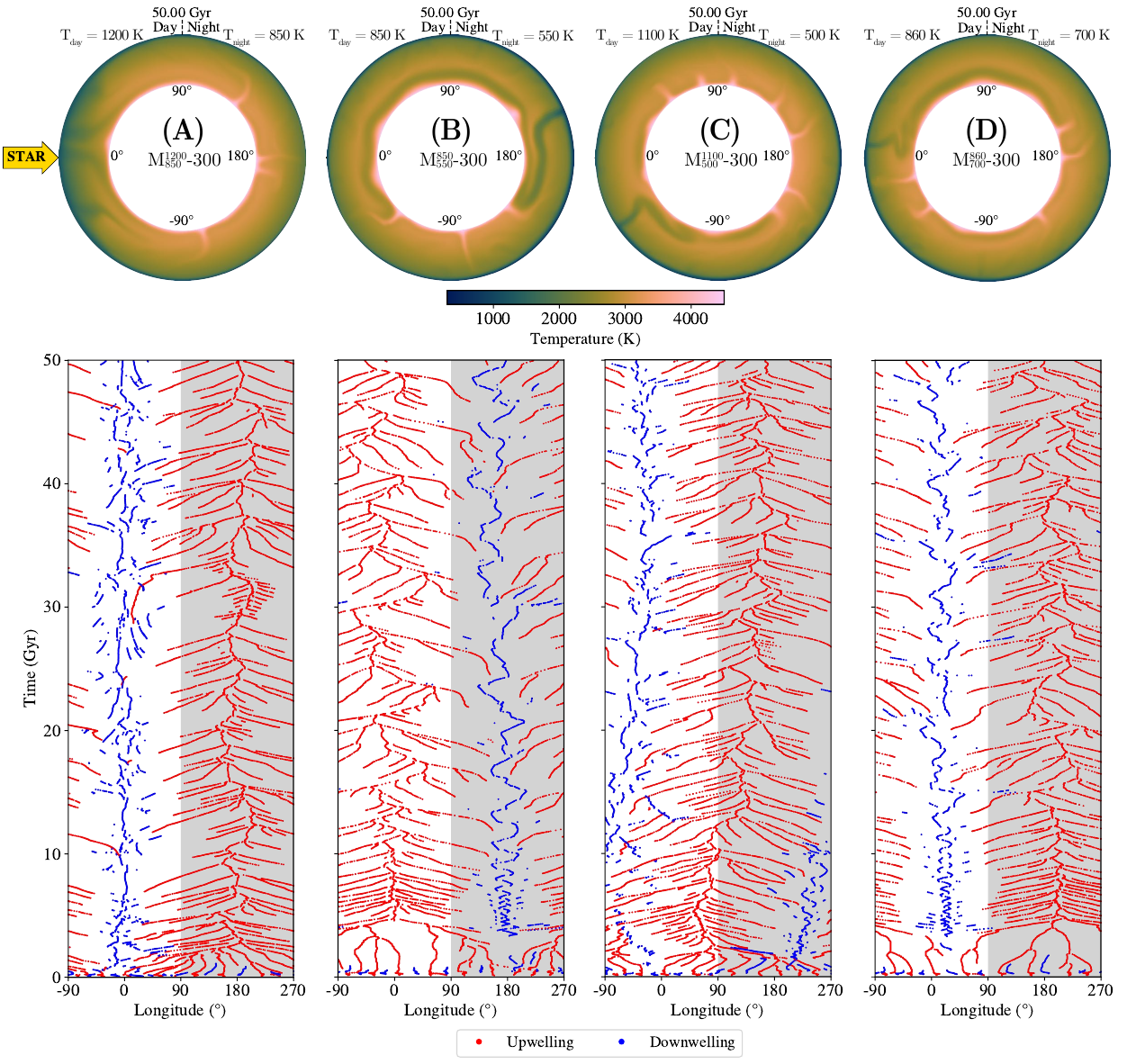}
    \caption{Same as Figure~\ref{fig:gj_theo_tzebra} but with surface temperatures that are derived from general circulation models. Each column (A--D) corresponds to a different surface temperature contrast. Similar to the models shown in Figure~\ref{fig:gj_theo_tzebra}, a degree-1 convection pattern is observed in all models, characterised by a strong downwelling forming on one hemisphere and upwellings being pushed towards the other side.}
    \label{fig:gj_gcm_tzebra}
\end{figure*}

Figures~\ref{fig:gj_theo_tzebra} and \ref{fig:gj_gcm_tzebra} show snapshots of the mantle temperature and evolutionary tracks of upwellings and downwellings for the models with a strong yield stress ($\sigma_{\text{duct}}=300$\,MPa) where the surface temperature were derived assuming different efficiencies of heat redistribution and from GCMs respectively.
For the model without heat redistrubition (M$^{990}_{10}$-300), a strong downwelling forms on the dayside of the planet and upwellings get pushed preferentially towards the nightside (Fig.~\ref{fig:gj_theo_tzebra}A). The nightside is more yielded than the dayside and is very viscous (as it reaches the maximum viscosity allowed by the numerical model). Initially, a downwelling forms closer to the terminator at the nightside, but later breaks away. 
Assuming average heat redistribution (M$^{845}_{350}$-300), a downwelling forms at the top terminator where it oscillates over time between the dayside and the nightside (Fig.~\ref{fig:gj_theo_tzebra}B). Upwellings are preferentially around the lower terminator. The nightside upper mantle is rather viscous, although the thermal boundary layer is not as thick as for the M$^{990}_{10}$-300 model. Before the downwelling forms, cold material accumulates around the upper terminator. An upwelling at the substellar point leads to strong deformation and strain at the surface and the upwelling then moves towards the terminator. At the terminator, cold material will then subduct into the deep mantle. 
For the models with a uniform surface temperature, a degree-1 convection pattern is also established for both cases. For model with a hotter surface temperature, (M$^{700}_{700}$-300) a downwelling forms and moves around the terminator (Fig.~\ref{fig:gj_theo_tzebra}C). Towards the end of the run, the downwelling has moved approximately $45^{\circ}$ towards the anti-stellar point. For the model with an Earth-like surface temperature, the downwelling forms around the terminator and then also oscillates between the substellar point and terminator region (Fig.~\ref{fig:gj_theo_tzebra}D).   

Model M$^{1200}_{850}$-300 displays a similar convection pattern to M$^{990}_{10}$-300: A strong downwelling forms on the dayside and upwellings are getting pushed towards the nightside. A return flow of cold material from the nightside towards the dayside replenishes the cold thermal boundary layer on the dayside before it accumulates enough buoyancy and breaks away (Fig.~\ref{fig:gj_theo_tzebra}A). This return flow of cold material is accommodated in the upper mantle. 
Similarly for model M$^{860}_{700}$-300, prominent downwelling forms on the dayside and upwellings get pushed towards the nightside. Compared to the M$^{990}_{10}$-300 model, the cold thermal blanket that occasionally forms on the dayside, is less thick (\ref{fig:gj_gcm_tzebra}D). The temperature contrast for this model is relatively low, and consequently there are no pronounced differences in viscosity, strain rate or yielding between the dayside and nightside.
Model M$^{1100}_{500}$-300 is also characterised by a prominent downwelling on the dayside and upwellings are located on the nightside. Up to $12$\,Gyr, the downwelling is on the nightside (close to the terminator) and slowly moves towards the dayside (\ref{fig:gj_gcm_tzebra}C). On the dayside, smaller downwellings are forming occasionally and then merge with the already present strong downwelling over the whole course of the model run.

\subsection{Stability Analysis of Convection Pattern}
\begin{figure*}
    \centering
    \includegraphics[width=\linewidth]{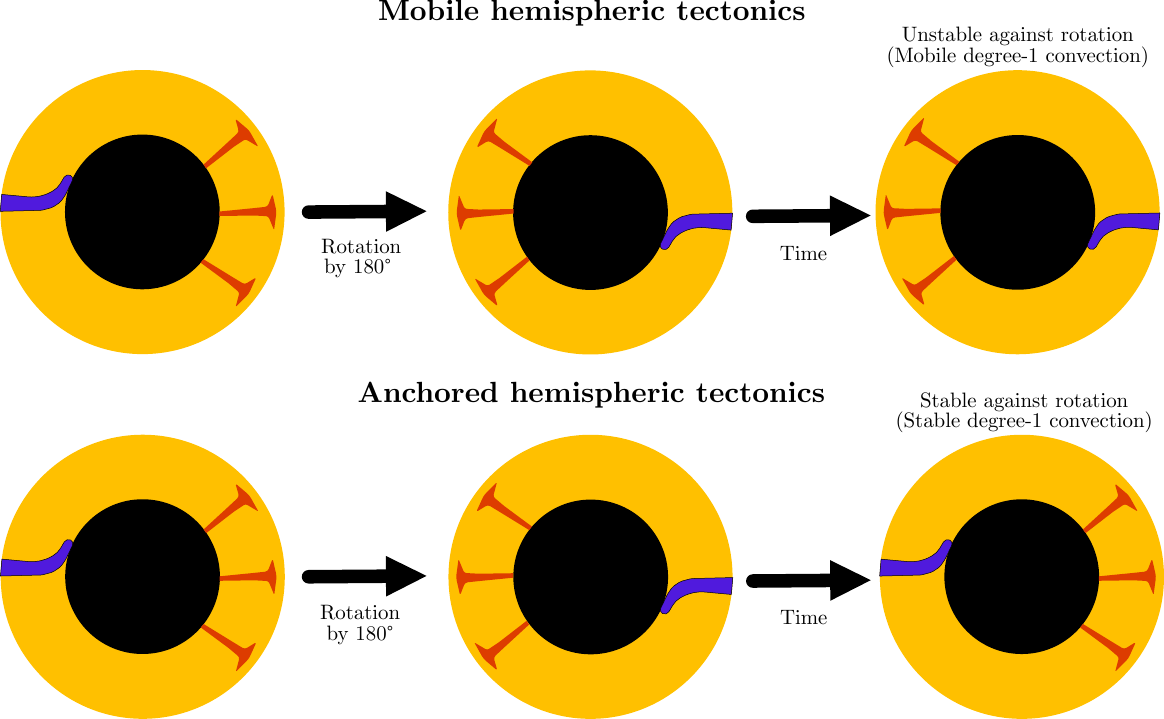}
    \caption{Illustration of rotated model runs. The initial state of the rotated model run corresponds to the $180^{\circ}$-rotated state where a degree-1 convection pattern has already been established. If the accompanying steady-state corresponds to the original, non-rotated state, we call the convection pattern stable degree-1 convection (the model is stable against the rotation). If the rotated state is also a stable convection regime, we call this a mobile degree-1 convection regime (the model is unstable against the rotation).}
    \label{fig:rotation_illustration}
\end{figure*}

\begin{figure*}
    \centering
    \includegraphics[width=\linewidth]{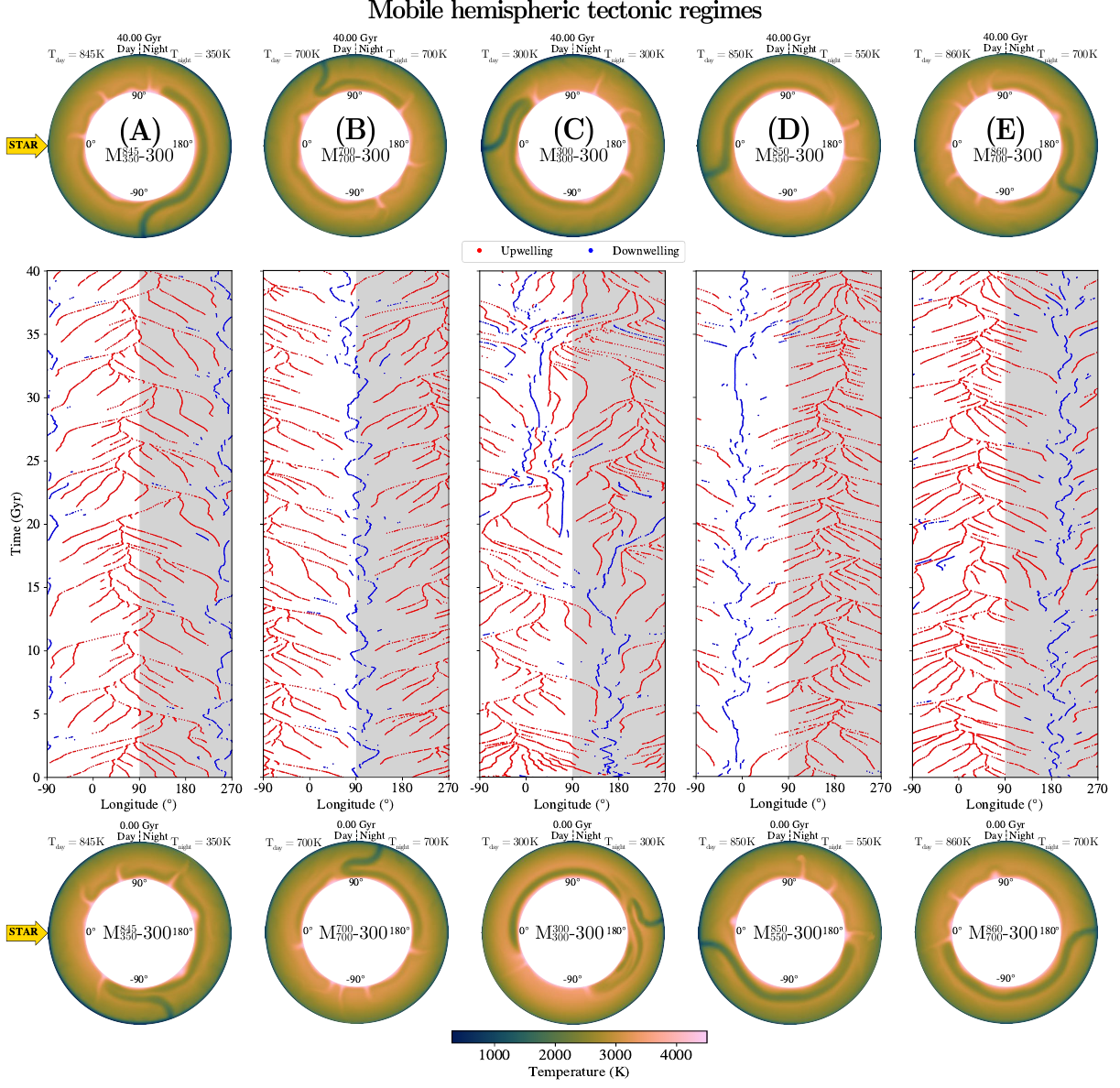}
    \caption{Snapshots of mantle temperature for super-Earth GJ~486b at $0$ (bottom row) and $40$\,Gyr (top row) and the corresponding evolutionary tracks of upwellings (red) and downwellings (blue) (middle row) for the models with a strong lithosphere (\mbox{$\sigma_{\text{duct}}=300$\,MPa}) for which a mobile degree-1 convection pattern is established. The initial mantle temperature (bottom row) corresponds to the $180^{\circ}$-rotated end-of-run state of the models shown in Figs.~\ref{fig:gj_theo_tzebra} and \ref{fig:gj_gcm_tzebra}. The substellar point is at $0\degree$ longitude (indicated by the yellow arrow) and the nightside ($90\degree$--$270\degree$) is denoted by a grey background. Each column (A--E) corresponds to a different surface temperature contrast. Mobile degree-1 convection is linked to to mobile hemispheric tectonics. Videos for these model runs are available on Zenodo \cite{MeierZenodo2024}.}
    \label{fig:gj_rot_mobile}
\end{figure*}

\begin{figure*}
    \centering
    \includegraphics[width=0.9\linewidth]{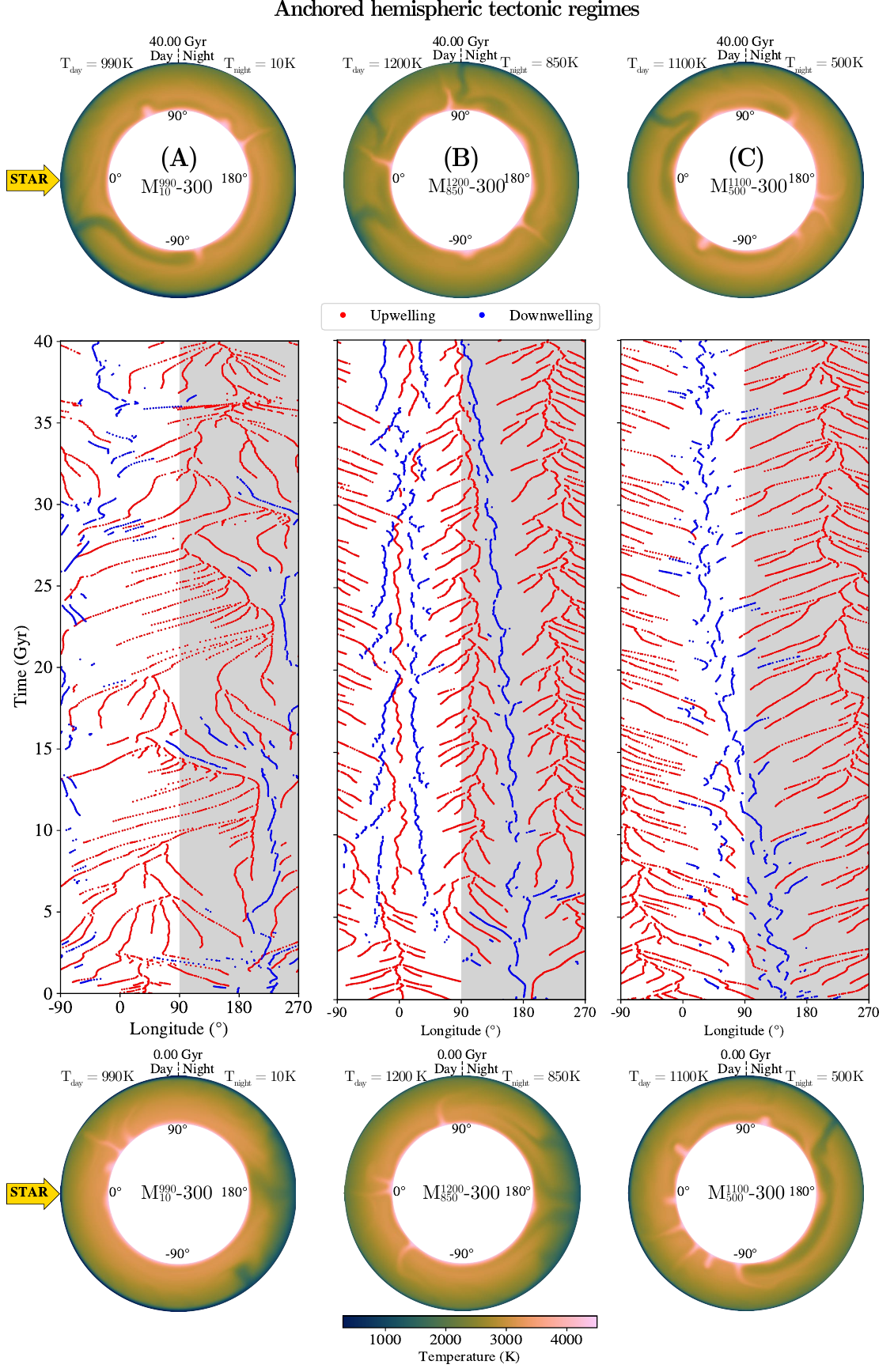}
    \caption{Same as Figure~\ref{fig:gj_rot_mobile} but for the models for which a stable degree-1 convection pattern is established. The initial mantle temperature (bottom row) corresponds to the $180^{\circ}$-rotated end-of-run state of the models shown in Figs.~\ref{fig:gj_theo_tzebra} and \ref{fig:gj_gcm_tzebra}. Each column (A--C) corresponds to a different surface temperature contrast. As can be seen from the evolutionary plots (middle row), donwellings (blue) or upwellings (red) now migrate towards the location of the non-rotated end-of-run state. Stable degree-1 convection is linked to anchored hemispheric tectonics. Videos for these model runs are available on Zenodo \cite{MeierZenodo2024}.}
    \label{fig:gj_rot_anchored}
\end{figure*}

Because all models with a strong lithosphere ($\sigma_{\text{duct}}=300$\,MPa) develop a degree-1 convection pattern regardless of surface temperature contrasts, we now investigate the stability of this convection pattern. In order to analyse the effect of the surface temperature contrast, we ran `rotated models' using an initial state that corresponds to the $180^{\circ}$-rotated state where a degree-1 convection pattern has already been established. Figure~\ref{fig:rotation_illustration} illustrates how these rotated models are used to distinguish between a stable and mobile degree-1 convection pattern. A stable degree-1 convection is linked to anchored hemispheric tectonics whereas mobile degree-1 convection is linked to mobile hemispheric tectonics. 
Figures~\ref{fig:gj_rot_mobile} and \ref{fig:gj_rot_anchored} show the models which lead to mobile and anchored hemispheric tectonics respectively. Models M$^{845}_{350}$, M$^{700}_{700}$, M$^{300}_{300}$, M$^{850}_{550}$, and M$^{860}_{700}$ are all in a mobile degree-1 convection regime. The evolutionary plots show that the rotated configurations are also stable. This is particularly apparent for the M$^{845}_{350}$, M$^{700}_{700}$, M$^{850}_{550}$, and M$^{860}_{700}$ models where the downwelling is stable in the rotated position for several tens of Gyrs (Figs~\ref{fig:gj_rot_mobile}: A, B, D, E). The M$^{300}_{300}$ model is stable for approximately $20$\,Gyr after which it completely subducts into the deep mantle. A new downwelling then forms around the terminator moving towards the dayside. This configuration is then stable for the next $20$\,Gyr. Both a downwelling on the nightside and dayside is therefore stable, and hence this model is in a mobile hemispheric tectonic regimes, as expected from the uniform surface temperature contrast. 
Models M$^{990}_{10}$, M$^{1200}_{850}$, M$^{1100}_{500}$ display a stable degree-1 convection pattern. The rotated model configurations are not stable in these cases. 
In the case of M$^{1200}_{850}$ the downwelling, now initially on the nightside gradually shifts back towards the dayside over an extended period of approximately $40$\,Gyr. New downwellings form on the dayside after $\approx$\,$5$\,Gyr (Fig.~\ref{fig:gj_rot_anchored}B). 
Similarly for M$^{1100}_{500}$, the downwelling has moved back to the dayside after approximately $10$Gyr and upwellings are again preferentially on the nightside (Fig.~\ref{fig:gj_rot_anchored}C). For  M$^{990}_{10}$, the downwelling is rather stable on the nightside too, but is again preferentially on the dayside after $35$\,Gyr. Upwellings have a clearer inclination to move back towards the nightside (Fig.~\ref{fig:gj_rot_anchored}A).

\subsection{Varying the Yield Stress}

\begin{figure*}
    \centering
    \includegraphics[width=0.9\linewidth]{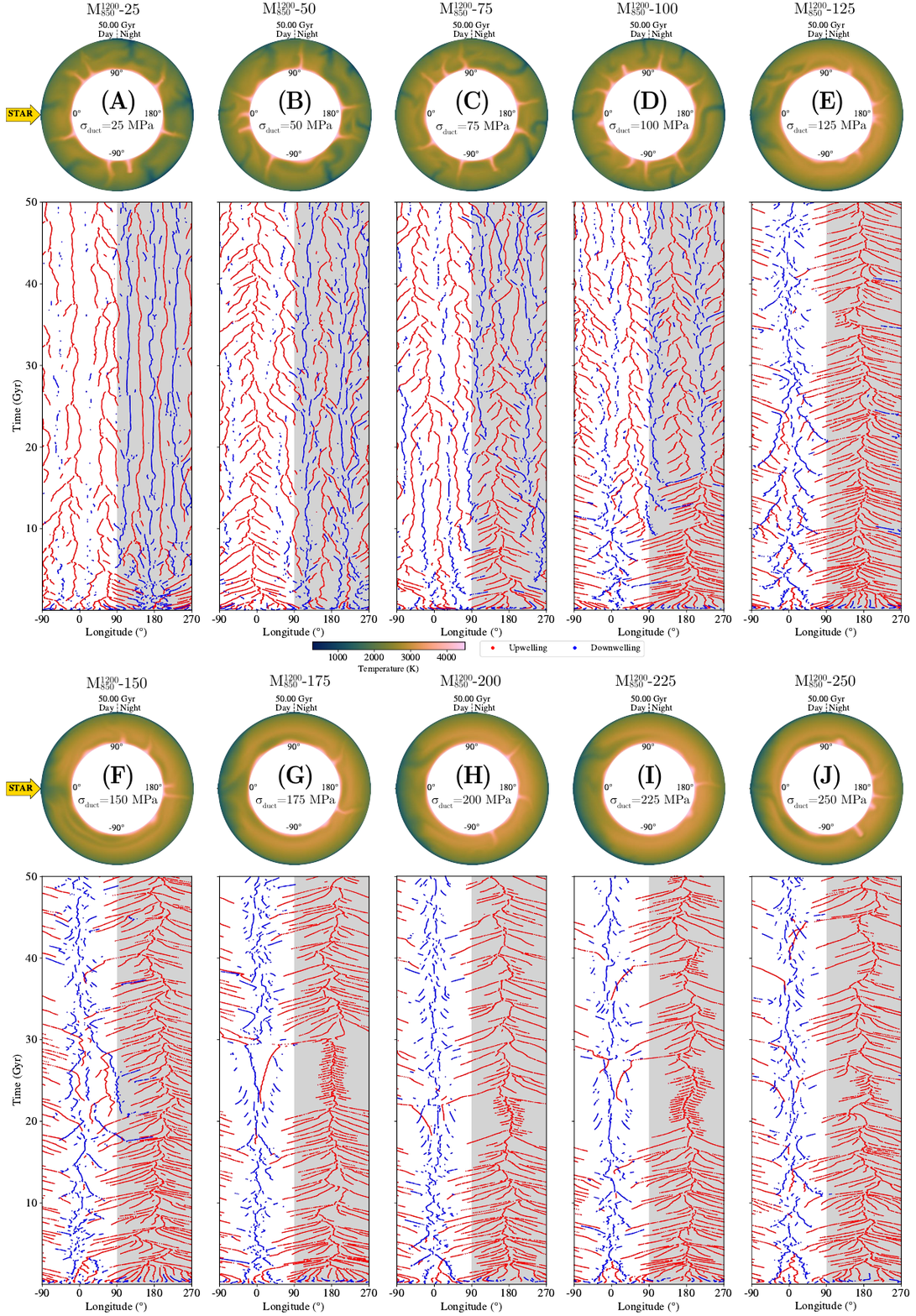}
    \caption[Mantle temperature and evolutionary plots for different strengths of the lithosphere (M$^{1200}_{850}$)]{Snapshots of mantle temperature (top panels) and the corresponding evolutionary tracks of upwellings (red) and downwellings (blue) (lower panels) for the M$^{1200}_{850}$ models for different strengths of the lithosphere ($\sigma_{\text{duct}}$). Each column (A--J) corresponds to a different strength of the lithosphere. The substellar point is at $0\degree$ longitude (indicated by the yellow arrow) and the nightside ($90\degree$--$270\degree$) is denoted by a grey background. A degree-1 convection pattern is established for $\sigma_{\text{duct}} \geq 125$\,MPa.}
    \label{fig:gj_1200-850_sigma}     
\end{figure*}

Figure~\ref{fig:gj_1200-850_sigma} shows snapshots of the mantle temperature and evolutionary tracks of upwellings and downwellings for the M$^{1200}_{850}$ model for varying strengths of the lithosphere. $\sigma_{\text{duct}}$ was varied from $25$\,MPa to $250$\,MPa. For $\sigma_{\text{duct}} \geq 125$\,MPa, a degree-1 convection pattern forms with a strong downwelling preferentially on the dayside and upwellings that are getting pushed towards the nightside. For very low yield stress ($\sigma_{\text{duct}} \leq 50$\,MPa), a uniform convection pattern is established: upwellings and downwellings are distributed uniformly around the mantle with stronger downwellings on the nightside.  For medium yield stress ($75$\,MPa $\leq \sigma_{\text{duct}} \leq 100$\,MPa), the convection pattern is still uniform, and the downwellings become equally strong on both sides (i.e., they have weakened on the nightside compared to the models with lower yield stress). The transition regime between a uniform and hemispheric tectonic regime is occurring between $100$\,MPa $\leq \sigma_{\text{duct}} \leq 125$\,MPa. In this regime, downwellings are preferentially on the dayside and they do not develop anymore on the nightside. Upwellings are still uniformly distributed around the CMB and downwellings on the dayside are of smaller scale than in the hemispheric tectonic regime.

\begin{figure*}
    \centering
    \includegraphics[width=0.9\linewidth]{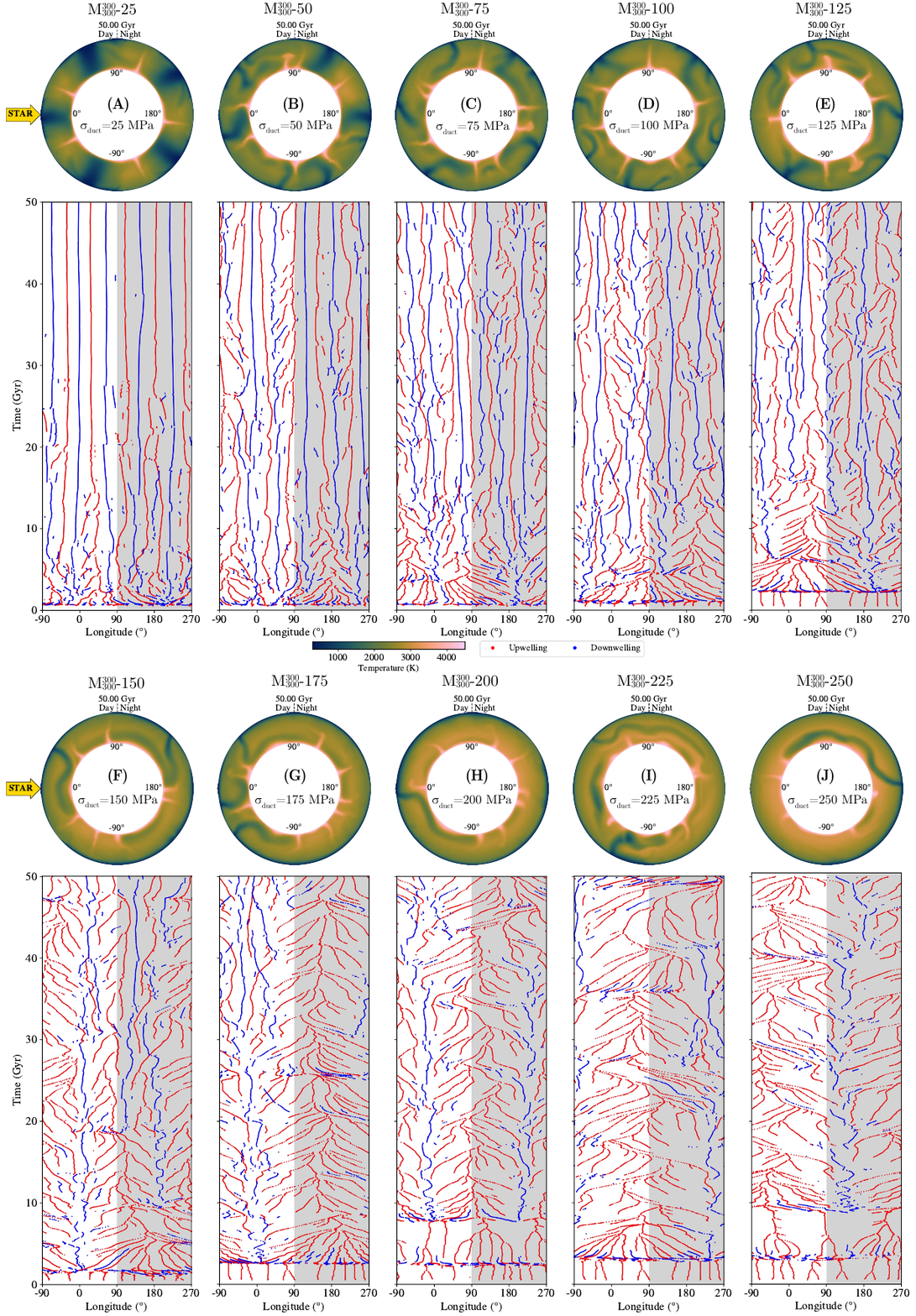}
    \caption{Same as Figure~\ref{fig:gj_1200-850_sigma} but with a uniform surface temperature of $300$\,K. For lower yield stress, upwellings and downwellings are uniformly distributed. Each column (A--J) corresponds to a different strength of the lithosphere. A degree-1 convection pattern is established for $\sigma_{\text{duct}} \geq 250$\,MPa.}
    \label{fig:gj_300-300_sigma}     
\end{figure*}

Figure~\ref{fig:gj_300-300_sigma} shows snapshots of the mantle temperature and evolutionary tracks of upwellings and downwellings for the M$^{300}_{300}$ model for varying strengths of the lithosphere $\sigma_{\text{duct}}$, which was varied from $25$\,MPa to $250$\,MPa.
For lower yield stresses ($\sigma_{\text{duct}} \leq 125$\,MPa), upwellings and downwellings are uniformly distributed. The downwellings are strongest for the models with very low yield stress and become fainter with increasing yield stress. A degree-1 convection pattern is established for $\sigma_{\text{duct}} \geq 250$\,MPa (the model with $\sigma_{\text{duct}}=300$\,MPa is shown in Fig.~\ref{fig:gj_theo_tzebra}D). The transition between a uniform convection pattern and a degree-1 convection pattern happens between $150$\,MPa $\leq \sigma_{\text{duct}}\leq 225$\,MPa. Model M$^{300}_{300}$-200 has a degree-1 convection pattern after $50$\,Gyr, but it is degree-2 (two strong downwellings, one at the substellar and one at the anti-substellar point) for most of the model run (Fig.~\ref{fig:gj_300-300_sigma}H).  
For increased yield stress there is also a prolonged period before downwellings begin to form. In the case of a lower yield stress ($<100$\,MPa), downwellings develop almost instantly whereas for a higher yield stress the cold upper thermal boundary layer first needs to build up before downwellings develop. Therefore, the mantle is first dominated by hot upwellings rising from the CMB towards the surface. 

\section{Discussion} \label{sec:discuss}

\subsection{Convection Patterns} \label{section:discussion_degree1} 

\begin{figure*}
    \centering
    \includegraphics[width=\linewidth]{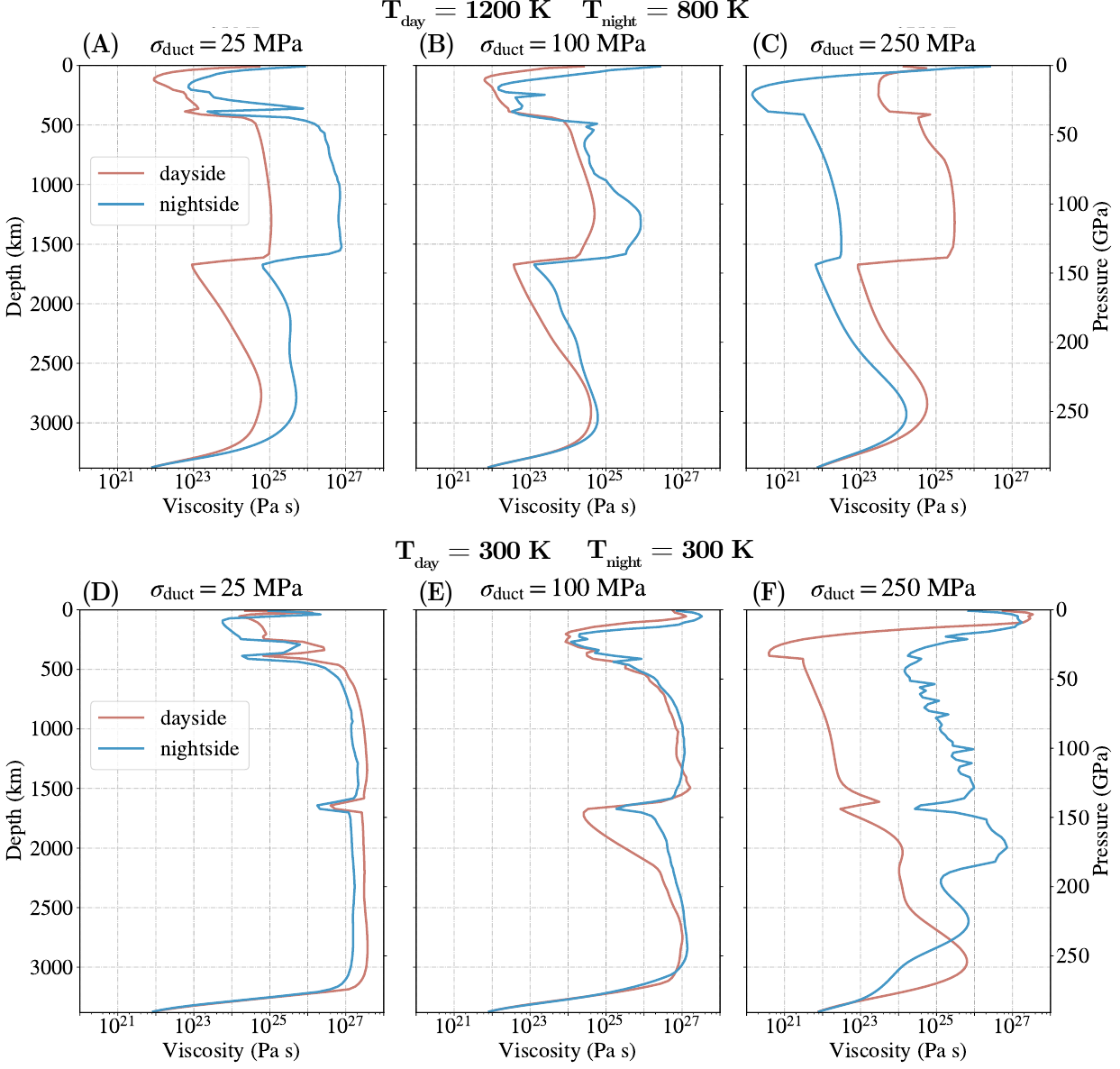}
    \caption{Horizontally averaged viscosity profiles for the models with \mbox{$T_{\textrm{day}}=1200$\,K} and \mbox{$T_{\textrm{night}}=850$\,K} (top row), and models with a uniform surface temperature of $300$\,K (bottom row). Each panel (A--J) corresponds to a different strength of the lithosphere. For both surface temperature contrasts, a degree-1 convection pattern is established for \mbox{$\sigma_{\text{duct}}=250$\,MPa}. With higher yield stress, the boundary layer has more time to grow, resulting in an increased viscosity contrast.}
    \label{fig:viscosity_profiles}     
\end{figure*}

Our mantle convection models of super-Earth GJ~486b show that both the strength of the lithosphere and the surface temperature contrast between the dayside and nightside have a strong control over the convection pattern. For the models with a weak lithosphere we find a similar convection pattern as we found for LHS\,3844b, which is a uniform distribution of upwellings and downwellings (Fig.~\ref{fig:gj_tplot_theo_weak}). GJ~486b has the same radius within generous error bars as LHS\,3844b ($1.3R_{\oplus}$), for which we found a uniform distribution of upwellings and plumes for cases with a weak lithosphere and hemispheric tectonics for high yield stresses (strong lithosphere) \cite{Meier2021}. 
Here, we find that increasing the ductile yield stress (strength of the lithosphere) increases the wavelength of mantle flow, which is consistent with previous studies that investigated the influence of yield strass on the convection pattern \cite{Tackley2000, Heck2008, Mallard2016}. A degree-1 convection pattern is established for all models of GJ~486b with high yield stress (Figs.~\ref{fig:gj_1200-850_sigma} \& \ref{fig:gj_300-300_sigma}) regardless of the temperature contrast between the dayside and nightside. This is most likely due to a stronger viscosity contrast across the top thermal boundary layer for larger yield stresses. Figure \ref{fig:viscosity_profiles} shows the horizontally averaged viscosity profiles for selected cases of the M$^{1200}_{850}$ and M$^{300}_{300}$ models. For larger yield stresses, the boundary layer has more time to grow, which increases the viscosity contrast through temperature and pressure dependence (Figs.~\ref{fig:viscosity_profiles}: C, F). If the yield stress is low, smaller stresses are needed to yield the lithosphere creating an almost iso-viscous layer (where $\eta_{\mathrm{eff}}$ is given by Equation~\eqref{chap3:eta_eff}). This creates a steady state with downwellings and upwellings that are almost immobile (Fig.~\ref{fig:gj_300-300_sigma}A and Figs~\ref{fig:viscosity_profiles}: A, D). By increasing the yield stress, the model becomes less steady as the increasing yield stress adds more non-linearity to the model and therefore also increases the time dependence. 
The models with high yield stress also feature a weaker upper mantle, with viscosity up to $3$ orders of magnitude lower than in models with a lower yield stress. This is consistent with previous studies which have shown that a strong viscosity contrast between the rigid lithosphere and a weaker upper mantle can lead to degree-1 convection \cite{Roberts2006, Liu2015, Zhong2016}. Previous studies have also shown that temperature and pressure dependence of viscosity will cause long-wavelength structures of convection \cite{Hansen1993, Bunge1996, Tackley1996, Tackley1996a,Yoshida2006, McNamara2005}. 
In this study, all models are in a mobile lid regime as we did not consider any stagnant lid cases. A stagnant lid regime could be recovered using a very high yield stress ($\sigma_{\mathrm{duct}} \to \infty$). \citeA{Meier2021} previously analysed this for super-Earth LHS\,3844b and found that this leads to no interior dichotomy with weak upwellings that are uniformly distributed uniformly and no downwellings. Degree-1 convection could however still operate below a stagnant lid if a weak asthenosphere or upper mantle is present below the rigid lithosphere \cite{Zhong2001, Roberts2006}. Such a weakening could for example result from partial melting in the shallower mantle \cite<e.g.>[]{Anderson1970, Karato2012}.

Our models for LHS\,3844b assumed an Earth-like core-to-radius ratio ($\approx$\,$0.55$), whereas for GJ~486b we used a core-to-radius ratio of $0.59$, which we determined by solving the internal structure equations (Section~\ref{sec:model}). 
\citeA{Caballero2022} used a Bayesian interference method to determine the core-to-radius ratio and found a value ranging from $0.40$--$0.46$. 
A smaller core size could make a degree-1 convection pattern more likely \cite{Zhong2000}. Similarly, including internal heating would likely increase the likelihood for degree-1 convection \cite{McNamara2005}.  Including internal heating also increases the mantle temperature, which in turn decreases the viscosity. This reduction in viscosity shortens the overturn time scales, as shown in \citeA{Meier2021}. For LHS\,3844b, \cite{Meier2021} found that including internal heating pushes the models with lower yield stress into a hemispheric tectonic regime, while the pattern of flow is steady and uniform if no internal heating is included. The overturn time scales are therefore most significantly influenced by the viscosity parameters, which could be explored by a future study.   
For this study, we assumed an Earth-like mineralogy and corresponding equation of state (Table~\ref{tab:density_refstate}). Even though the planet's bulk density can be inferred from its radius and mass measurements, there still remains a degeneracy in terms of its composition and interior structure \cite<e.g.>[]{Dorn2015, Spaargaren2023, Guimond2024}. A future study could therefore investigate how different mantle mineralogies and core sizes affects the geodynamic regimes of rocky super-Earths such as GJ~486b.

\subsection{Stability of Convection Pattern}  \label{section:stability_convection}
In this section, we address the influence of the surface temperature contrast on the stability of the tectonic regime and discuss how a strong temperature contrast between the dayside and nightside is necessary to anchor the downwelling to the nightside leading to anchored hemispheric tectonics. The models with a strong lithosphere ($\sigma_{\text{duct}}=300$\,MPa) all develop a degree-1 convection pattern (Figs.~\ref{fig:gj_theo_tzebra} and \ref{fig:gj_gcm_tzebra}). This pattern also forms for moderate temperature contrasts or even no temperature contrast between the dayside and nightside (M$^{300}_{300}$-300, M$^{700}_{700}$-300, Fig.~\ref{fig:gj_theo_tzebra}: C, D). This indicates that the degree-1 convection pattern is a consequence of rheology---particularly the strength of the lithosphere---rather than the temperature contrast between the dayside and nightside.

\begin{figure*}
    \centering
    \includegraphics[width=1.0\textwidth]{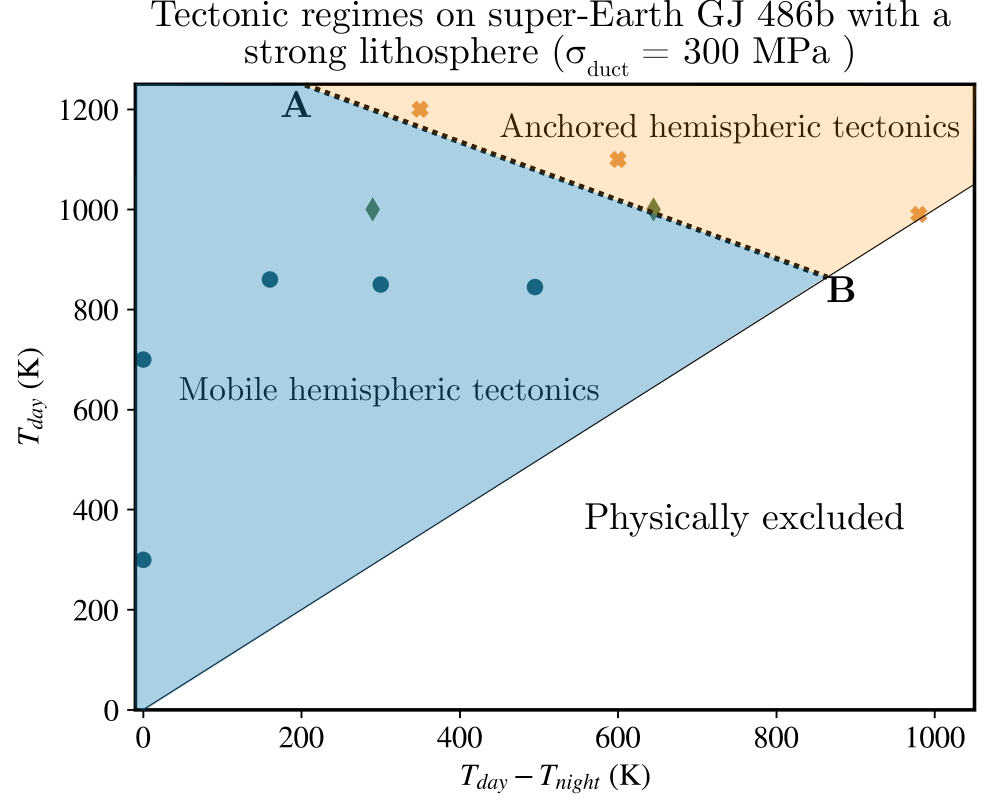}
    \caption[Overview of the different tectonic regimes]{Overview of the tectonic regimes for the models with a strong lithosphere. An increased temperature contrast between the dayside and nightside as well as an increased (dayside) temperature increases the likelihood of anchored hemispheric tectonics (orange shaded region). For lower temperature contrasts and lower dayside temperatures, the mantle is in a mobile degree-1 convection regime with no preferred location for upwellings and downwellings (blue shaded region). The diamond markers indicate models from our study on LHS\,3844b. The choice for the points A and B is explained in Section \ref{section:stability_convection}. More model runs are necessary to determine where the transition from mobile to anchored hemispheric tectonics is. The white region is excluded as this would result in negative temperature on the nightside.}
    \label{fig:gliese_summary_figure}
\end{figure*}

Our rotated model runs (for which the initial state that corresponds to the $180^{\circ}$-rotated state where a degree-1 convection pattern has already been established) indicate that the surface temperature contrast between the dayside and nightside controls whether this degree-1 convection pattern is stable or unstable. In a stable degree-1 convection regime, downwellings are preferentially on one hemisphere and upwellings are getting pushed towards the other side, whereas in a mobile degree-1 convection regime downwellings and upwellings do not have a preferred location. Stable degree-1 convection is linked to anchored hemispheric tectonics, whereas mobile degree-1 convection is linked to mobile hemispheric tectonics.   
Model M$^{990}_{10}$-300 shows a similar hemispheric tectonic regime as for the strong lithosphere case of LHS\,3844b \cite{Meier2021}, which is a $1.3\,R_{\oplus}$ super-Earth that supposedly does not have an atmosphere \cite{Kreidberg2019}. Since both planets have the same radius within error bars, we therefore expect that the tectonic regime of GJ~486b with no atmosphere is similar to that of LHS\,3844b. For LHS\,3844b with a strong lithosphere, we found that the strong surface temperature contrast can lead to hemispheric tectonics with a downwelling preferentially on the dayside and upwellings that are getting pushed towards the nightside. Since M$^{990}_{10}$-300 is in an anchored hemispheric tectonic regime, we expect super-Earth LHS\,3844b with a strong lithosphere to be in an anchored hemispheric tectonic regime too. Similar to the study of LHS\,3844b, we observe that in the stable degree-1 convection regimes of super-Earth GJ~486b, the downwelling is preferentially anchored to the dayside hemisphere, while upwellings are pushed towards the nightside. The reason for downwellings to be pushed towards the dayside rather than the nightside is most likely due to the temperature dependence of viscosity. Higher surface temperatures decreases viscosity, thereby facilitating mantle flow, with downwellings following the path of least resistance. Due to their high viscosity, downwellings will dictate the mantle flow pattern, pushing upwellings towards the opposite hemisphere. 
\begin{table*}[h!]
\caption{Overview of the tectonic regimes for the models with a strong lithosphere.$^{a}$}
\label{tab:rotated_models}
\centering
\begin{tabular}{ccccc}
\hline\hline
Model & dayside temp. & nightside temp.   & $T_{day}$ - $T_{night}$ & tectonic regime \\ 
& (K) & (K)  &  (K) &  \\
\hline 
M$^{990}_{10}$-300    &   990   & 10   &  980 & anchored hemispheric  \\ 
M$^{845}_{350}$-300   &   845   & 350  &  495 & mobile hemispheric \\ 
M$^{700}_{700}$-300   &   700   & 700  &  0   & mobile hemispheric \\ 
M$^{300}_{300}$-300   &   300   & 300  &  0   & mobile hemispheric\\ 
M$^{1200}_{850}-$300  &   1200  & 850  &  350 & anchored hemispheric\\ 
M$^{850}_{550}-$300   &   850   & 550  &  300 & mobile hemispheric\\
M$^{1100}_{500}-$300  &   1100  & 500  &  600 & anchored hemispheric \\ 
M$^{860}_{700}-$300   &   860   & 700  &  160 & mobile hemispheric  \\ 
\hline
\multicolumn{5}{p{1.0\linewidth}}{$^{a}$ If the tectonic regime is anchored, this indicates that downwellings move back towards their un-rotated position. If the tectonic regime is mobile, the downwelling will also be stable at the rotated position. This indicates a degree-1 convection pattern, where the downwellings and upwellings do not have a preferred location.}
\end{tabular}
\end{table*}
Table~\ref{tab:rotated_models} shows an overview of the different models and whether their tectonic regime is mobile hemispheric tectonics or anchored hemispheric tectonics.
Figure~\ref{fig:gliese_summary_figure} shows an overview of the tectonic regimes for GJ~486b for the case of a strong lithosphere. 
Overall, a strong temperature contrast between the dayside and nightside favours an anchored hemispheric tectonic regime. Also, the dayside temperature needs to be sufficiently hot for anchored hemispheric tectonics, as can be seen from model  M$^{845}_{300}$, which has a rather strong temperature contrast but does not develop anchored hemispheric tectonics. 
Point A in Figure~\ref{fig:gliese_summary_figure} has been defined assuming that anchored hemispheric tectonics requires a minimum temperature contrast of $\approx 200$\,K. For LHS\,3844b, we have also run models with nightside temperatures of $T_{\textrm{night}}=355$\,K and $710$\,K. Even though at the time of that study, we concluded that hemispheric tectonics persists even for hotter nightside temperatures, it is now apparent that the model with $T_{\textrm{night}}=710$\,K (with basal heating) actually exhibits a mobile hemispheric tectonic regime, and not anchored hemispheric tectonics. 

The model of LHS\,3844b with $T_{\textrm{night}}=355$\,K is rather stable as downwellings that move towards the dayside disappear quickly again \cite<>[Fig.~3B]{Meier2021}, but it is difficult to say whether this regime is actually anchored hemispheric tectonics without running a rotated model. We estimate the location of Point B in Figure~\ref{fig:gliese_summary_figure} by assuming that the LHS\,3844b model with $T_{\textrm{night}}=355$\,K and $T_{\textrm{day}}=1000$\,K lies at the boundary between mobile and anchored hemispheric tectonics. However, more model runs are necessary to define this transition from mobile to anchored hemispheric tectonics. We can assume, however, that planets with a very small temperature contrast do not exhibit anchored hemispheric tectonics and that this likelihood increases for planets with a very strong temperature contrast even if the dayside temperature is not very hot.  

As mentioned in Section~\ref{sec:tracking}, we ran the models for several tens of billions of years as we are looking for a steady flow solution and are interested in the statistical distribution of plumes and downwellings. Such long integration times are necessary due to the higher viscosity, resulting in more sluggish convection and longer overturn timescale compared, for example, Earth. Our findings show that in the cases of anchored hemispheric tectonics, where the downwelling gradually returns to a preferred location, this process can occur over an extended period, spanning several billions of years very long time, potentially exceeding the age of the system itself. A future study could therefore explore the influence of initial conditions and whether upwellings and downwellings are more or less likely to be located at their preferred, anchored locations early on in the geodynamic evolution. 

\subsection{Relevance to Observations}

As mentioned in Section~\ref{section:intro}, GJ~486b has recently been observed with JWST \cite{Moran2023} and high-resolution spectroscopy \cite{RiddenHarper2023}. However, it is still debated whether GJ~486b possesses an atmosphere or not, as the water feature in the JWST transmission spectrum could also be caused by water that is present in cold star spots contaminating the transmission spectrum \cite{Moran2023}. Transit spectroscopy is probing the day-night terminator region of an atmosphere \cite<e.g.>[]{Perryman2018, Kreidberg2018}. Some of our models show an asymmetry between the terminators with a downwelling oscillating around one terminator and upwellings getting pushed towards the opposite terminator. These models are in a mobile degree-1 convection regime but not anchored degree-1 convection as the temperature contrast is not driving downwellings and upwellings towards a preferred location. If upwellings have a preferred location (e.g., nightside or terminators) that is stable over geological timescales, this could lead to preferential melt generation and outgassing at that location \cite<e.g.>[]{Black2019}. 

Transit spectroscopy usually assumes that the terminator region is homogeneous. Space missions, such as JWST, TESS or CHEOPS could probe asymmetries between the morning and evening terminators \cite{Espinoza2021}. High-resolution transmission spectroscopy has been used to infer asymmetries between the morning and evening terminators of the ultra-hot Jupiter WASP-76b \cite{Ehrenreich2020, Wardenier2021, Kesseli2022} and WASP-121b \cite{AzevedoSilva2022}. For rocky exoplanets with smaller planet-to-star radius ratios, detecting such asymmetries is, however, much more challenging.
A north-south asymmetry would be more difficult to detect as the star light will pass through the northern and southern atmosphere almost instantaneously if the planet's orbital plane is not strongly inclined. North-south or morning-evening asymmetries arise due to the 3D nature of the planet. Therefore, it would also be necessary to run 3D interior dynamic models in order to investigate whether hemispheric tectonics persists and whether there is a preference for either a north-south or morning-evening asymmetry. 
If upwellings lead to preferential outgassing of volatiles on one hemisphere, a promising approach to detect an anchored hemispheric tectonic regime is to identify potential variations in atmospheric compositions between the dayside and nightside due to this preferential outgassing. However, detecting such asymmetries, which depend on the efficiency of heat redistribution and strength of localised outgassing, is likely beyond the capabilities of current telescopes, such as JWST. Consequently, future studies are needed that quantify the amount of melt production and outgassing rates on tidally-locked super-Earths. 

Emission spectroscopy probes the dayside temperature structure and chemical composition of a planet's dayside atmosphere \cite<e.g.>[]{Burrows2014, Stevenson2014, Mansfield2023}. A reflection spectrum in the mid- to near-infrared can be used to identify different rocky surface types of an exoplanet \cite{Hu2012, Whittaker2022, Byrne2023}.
The nightside of a planet remains difficult to probe. Phase resolved emission spectra can give insights into the thermal structure of an exoplanet's atmosphere \cite<e.g.>[]{Stevenson2014}. Emission from a planet's nightside can usually be neglected at visible wavelengths, but could contribute to planet's transit depth at infrared wavelengths \cite{Kipping2010}. For super-Earth LHS\,3844b, \citeA{Meier2021} found that the contribution to the thermal phase curve from the interior flux alone would produce temperature differences on the nightside on the order of $15$--$30$\,K, which is significantly below the detection capabilities of current and near-future observations. The same conclusion applies to super-Earth GJ~486b as the flux from the interior is on the same order of magnitude. If upwellings on the nightside indeed lead to preferential melt generation on the nightside, these could imprint themselves in phase curve observations or attenuate the observed transit depth. 
In summary, a combination of different observation techniques (i.e., transmission spectroscopy, emission and reflection spectroscopy and thermal phase curve observations) will be necessary to confirm mobile or anchored hemispheric tectonics on a tidally locked super-Earth. Hence, consideration of the interior dynamics of rocky exoplanets such as GJ~486b will feed back into determining the requirements of future observational facilities and missions. 
   
\section{Conclusions} \label{sec:conc}
In this study, we investigated the different possible tectonic regimes of super-Earth GJ~486b using 2D numerical mantle convection models. We constrained the surface temperature contrasts assuming different heat efficiencies of atmospheric heat circulation and from general circulation models. We investigated how the mantle flow is affected by the corresponding surface temperature contrasts and how it depends on the strength of the lithosphere, which we modelled through employing a plastic yielding criterion. 

Models with a low yield stress ($\sigma_{\mathrm{duct}}=10$\,MPa) develop a uniform convection pattern with downwellings and upwellings uniformly distributed around the mantle. Downwellings are colder and more viscous on the nightside if the nightside is much cooler than the dayside. These regimes are similar to the tectonic regimes we found for super-Earth LHS\,3844b with a weak lithosphere \cite{Meier2021}. Models with a medium yield stress ($\sigma_{\mathrm{duct}}=100$\,MPa) develop a similar convection regime compared to the weak lithosphere cases, albeit downwellings and upwellings are more mobile. For models with a high yield stress ($\sigma_{\mathrm{duct}}=300$\,MPa), we find that a degree-1 convection pattern is established, independent of the temperature contrast between the dayside and nightside. An increased temperature contrast or overall higher surface temperature increases the likelihood of anchored hemispheric tectonics, in which the downwelling is preferentially pinned on one hemisphere. The transition from a uniform regime to degree-1 convection occurs around $\sigma_{\mathrm{duct}}=125$--$200$\,MPa depending on the surface temperature contrast. 

A suite of different observational techniques will be necessary in order to distinguish the different convection patterns that we find from our simulations. Similar to what we found for LHS\,3844b, the heat contribution from the interior will most likely not lead to significant spectral signatures in the thermal phase curve observations that could be detectable by current or near-future space observations. However, if upwellings lead to preferential melt generation and outgassing of volatiles on one hemisphere, this might lead to signatures that could be detected by transmission or emission and reflection spectroscopy.  Our results suggest that rocky super-Earths show highly variable tectonic regimes that are intrinsically linked to both thermal forcing from the star as well as compositional differences that affect interior properties.

%
%

\section*{Open Research Section}
Selected data for reproducing the figures in this study can be obtained from the Zenodo Data Repository \cite{MeierZenodo2024}. Post-processing of \textsc{StagYY} data was performed using \textsc{StagPy} \cite{Morison2022}, \textsc{numpy} \cite{Harris2020}, and \textsc{SciPy} \cite{Virtanen2020}. The figures were created using \textsc{matplotlib} \cite{Hunter2007}. The perpetually uniform colourmaps are from \citeA{Crameri2018}. The mantle convection code \textsc{StagYY} \cite{Tackley2008} is the property of P.J.T. and Eidgenössische Technische Hochschule Zürich (ETHZ) and is available for collaborative studies from Paul J. Tackley. Requests can be made to P.J.T. (\href{mailto:paul.tackley@erdw.ethz.ch}{paul.tackley@erdw.ethz.ch}).

\acknowledgments
The authors thank Shijie Zhong and an anonymous reviewer for taking the time to review this paper and for their constructive feedback, which has greatly helped improve the manuscript. T.G.M. and D.J.B. acknowledge SNSF Ambizione Grant 173992. T.G.M. was supported by the SNSF Postdoc Mobility Grant P500PT\_211044. Calculations were performed on UBELIX (\url{http://www.id.unibe.ch/hpc}) the HPC cluster at the University of Bern, and the AOPP HPC cluster. T.L. acknowledges support by the Branco Weiss Foundation, the Alfred P. Sloan Foundation (AEThER project, G202114194), and NASA’s Nexus for Exoplanet System Science research coordination network (Alien Earths project, 80NSSC21K0593). M.H. is supported by Christ Church, University of Oxford.
J.A.C. acknowledges financial support from the Agencia Estatal de Investigaci\'on (AEI/\allowbreak10.13039/501100011033) of the Ministerio de Ciencia e Innovaci\'on and the European Regional Development Fund ``A way of making Europe'' through project PID2022-137241NB-C42. This project received support from  STFC Consolidated Grant ST/W000903/1 and from the Alfred P. Sloan Foundation under grant G202114194 to the AETHER project. 
This research benefited from discussions and interactions within the framework of the National Center for Competence in Research (NCCR) PlanetS supported by the SNSF.

%
%

\end{document}